# Towards Deterministic Communications in 6G Networks: State of the Art, Open Challenges and the Way Forward

**GOURAV PRATEEK SHARMA[1], DHRUVIN PATEL[2], JOACHIM SACHS[3], MARILET DE ANDRADE[3], JANOS FARKAS[4], JANOS HARMATOS[4], BALAZS VARGA[4], HANS-PETER BERNHARD[5,6], RAHEEB MUZAFFAR[5], MAHIN K. ATIQ[5], FRANK DUERR[7], DIETMAR BRUCKNER[8], EDGARDO MONTESDEOCA[9], DRISSA HOUATRA[10], HONGWEI ZHANG[11] AND JAMES GROSS[1]**

[1] EECS School, KTH Royal Institute of Technology, Stockholm, Sweden
[2] Ericsson Research, Aachen, Germany
[3] Ericsson Research, Stockholm, Sweden
[4] Ericsson Research, Budapest, Hungary
[5] Silicon Austria Labs, Linz, Austria
[6] Johannes Kepler University Linz, Austria
[7] Institute for Parallel and Distributed Systems, University of Stuttgart, Germany
[8] B&R Industrial Automation, Eggelsberg, Austria
[9] Montimage, Paris, France
[10] Orange Innovation, Chatillon, France
[11] Iowa State University, Ames, USA

Corresponding author: Gourav Prateek Sharma (e-mail: gpsharma@kth.se).

This work was supported by the European Union's Horizon 2020 Research and Innovation Program under Grant 101096504 DETERMINISTIC6G (see deterministic6g.eu).

**Abstract** Over the last decade, society and industries are undergoing rapid digitization that is expected to lead to the evolution of the cyber-physical continuum. End-to-end deterministic communications infrastructure is the essential glue that will bridge the digital and physical worlds of the continuum. We describe the state of the art and open challenges with respect to contemporary deterministic communications and compute technologies— 3GPP 5G, IEEE Time-Sensitive Networking, IETF DetNet, OPC UA as well as edge computing. While these technologies represent significant technological advancements towards networking Cyber-Physical Systems (CPS), we argue in this paper that they rather represent a first generation of systems which are still limited in different dimensions. In contrast, realizing future deterministic communication systems requires, firstly, seamless *convergence* between these technologies and, secondly, *scalability* to support heterogeneous (time-varying requirements) arising from diverse CPS applications. In addition, future deterministic communication networks will have to provide such characteristics *end-to-end*, which for CPS refers to the entire communication and computation loop, from sensors to actuators. In this paper, we discuss the state of the art regarding the main challenges towards these goals: predictability, end-to-end technology integration, end-to-end security, and scalable vertical application interfacing. We then present our vision regarding viable approaches and technological enablers to overcome these four central challenges. In particular, we argue that there is currently a window of opportunity to establish, through 6G standardization, the foundations towards a scalable and converged deterministic communication and compute infrastructure. Key approaches to leverage in that regard are 6G system evolutions, wireless friendly integration of 6G into TSN and DetNet, novel end-to-end security approaches, efficient edge-cloud integrations, data-driven approaches for stochastic characterization and prediction, as well as leveraging digital twins towards system awareness.

**Index terms** 6G, URLLC, TSN, DetNet, wireless, machine learning, deterministic communication

## List of Abbreviation

The following is a list of abbreviations with their full forms that will be referred to later in the paper.

| | |
|---|---|
| 3GPP | 3rd Generation Partnership Project |
| AF | Application Function |
| AMF | Access and Mobility Management |
| API | Application Programming Interface |
| AVB | Audio-Video Bridging |
| CAPIF | Common API Framework |
| CBS | Credit-based Shaper |
| CG | Configured Grants |
| CNC | Centralized Network Configuration |
| CoMP | Coordinated Multipoint Transmission |
| CoMP | Coordinated Multipoint Transmission |
| CPS | Cyber-Physical Systems |
| CQF | Cyclic Queueing Forwarding |
| CUC | Centralized User Configuration |
| DetNet | Deterministic Networking |
| DoS | Denial of Service |
| DT | Digital Twins |
| eMBB | Enhancement Mobile Broadband |
| FLC | Field-Level Communications |
| FRER | Frame Replication and Elimination for Reliability |
| GM | Grandmaster |
| gNB | Next Generation NodeB |
| GNSS | Global Navigation Satellite System |
| gPTP | Generalized PTP |
| HARQ | Hybrid Automatic Repeat Request |
| IETF | Internet Engineering Task Force |
| ITU-T | International Telecommunication Union Telecommunication Standardization Sector |
| JCS | Joint Communication and Sensing |
| KPI | Key-Performance Indicator |
| LTE | Long-Term Evolution |
| MAC | Medium Access Layer |
| MCS | Modulation and Coding Scheme |
| ML | Machine Learning |
| NF | Network Function |
| NPN | Non-public Network |
| NSA | Non-Standalone |
| OE | Occupational Exoskeleton |
| OPC UA | Open Platform Communications Unified Architecture |
| PDU | Packet Data Unit |
| PDV | Packet Delay Variation |
| PLC | Programmable Logic Controller |
| PREOF | Packet Replication, Elimination and Ordering Function |
| PSFP | Per-stream Filtering and Policing |
| PTP | Precision Time Protocol |
| QoS | Quality of Service |
| RAN | Radio Access Network |
| RIS | Reconfigurable Intelligent Surface |
| RRC | Radio Resource Control |
| SBA | Service-based Architecture |
| SDN | Software Defined Networking |
| SDO | Standard Development Organization |
| SEAL | Service Enablement Architectural Layer |
| SGT | Secure Group Tags |
| SIB | System Information Block |
| SINR | Signal to Interference and Noise Ratio |
| SLAM | Simultaneous Localization and Sensing |
| SMF | Session Management Function |
| TDD | Time Division Duplexing |
| TSC | Time-Sensitive Communications |
| TSN | Time-Sensitive Networking |
| TT | TSN Translator |
| TTI | Transmission Time Interval |
| UE | User Equipment |
| UPF | User Plane Function |
| VN | Virtual Network |
| XR | Extended Reality |

## I. Introduction

Over the last decade, state-of-the-art in communications, networking and IT infrastructures has seen a steeply increasing interest in latency as - rediscovered - performance metric. In the wireless communications community, the pinnacle of this development has been the definition of Ultra-Reliable Low Latency Communications (URLLC), as one of three flavors of the 3GPP 5G mobile communications standard. In parallel, originating from tight latency requirements in Audio Video Bridging (AVB) applications, Time-Sensitive Networking (TSN) has been emerging as a major, unifying standard targeting Ethernet-based communications in vertical sectors such as industrial automation and manufacturing with stringent and deterministic latency constraints [1]–[5]. In a complementary initiative, the Internet Engineering Task Force (IETF) Deterministic Networking (DetNet) pursues similar goals but for L3-routed networks [6] and L4S for L3 queueing latency and congestion response [7]. Furthermore, with the advent of edge computing, the spatial proximity of public/open compute services translate into shortened access latencies, being one of the main advantages of edge computing over cloud computing [8]. Finally, latency needs to be addressed not only in L2 and L3, but also in transport and application layers L4-L7 and with regard to security.

These developments have to a large extent been triggered and motivated by use cases and stakeholders in the industrial automation and manufacturing vertical. Over the last 15 years, significant technical advances with respect to the design and operation of cyber-physical systems, robot systems as well as computer vision have incentivized the industry to renew digitalization efforts on the shopfloor, captured for instance in the concept of Industry 4.0 [9]. This, in turn, has enabled advanced manufacturing concepts like flexible automation, associated with a steadily increasing number of mobile systems on the shopfloor. 5G-based URLLC wireless communications, TSN, DetNet as well as edge computing systems cater to these needs. In addition, the automation community has been driving efforts to support shopfloor digitalization at scale through middleware standards like Open Platform Communications Unified Architecture (OPC UA). As of today, commercial systems based on all these standards have already found their way into industrial practice, mostly within small-scale scenarios embedded into controllable environments.

Nevertheless, as digitization is continuously driving the transformation of society and industries, the natural progression is a substantial acceleration towards the 2030 timeframe, as more and more technical enablers mature and spark innovations. Along this trajectory, entirely new forms of interactions will appear, between people, between people and machines, and in-between machines – all with the option of wireless communication. Furthermore, through cloud and edge computing, digital twins of physical entities and processes are enabled to create Cyber-Physical Systems (CPSs) that integrate physical and computational resources in order to control or monitor a mechanism. At the same time, artificial intelligence with data-driven system design opens novel dimensions for analytics and the optimizations of all sorts of processes. These trends are expected to result in a cyber-physical continuum, between the connected physical world of senses, actions, experiences, and its programmable digital representations [10]. This cyber-physical continuum is characterized by a massive scale-up and densification of CPSs such that increasingly these CPSs interact with each other at runtime. However, a prerequisite for this cyber-physical continuum is an intelligently networked infrastructure forming an efficient "glue" between the digital and physical worlds [11]. Currently, a unique window of opportunity has opened up to set the foundations of future 6G mobile networks to enable this emerging cyber-physical continuum. Two main goals need to be reached in this respect, going far beyond current standards and systems. Firstly, *convergence* among technologies, standards and infrastructures towards efficient provisioning of end-to-end properties for CPS applications will have to be achieved. However, going beyond this, the second goal to be achieved is *scalability* among operations and services of future deterministic infrastructures carrying CPS applications.

Let us start with a discussion about the related challenges with respect to the goal of *convergence*. Typical applications of the cyber-physical continuum will comprise CPSs such as control and automation, wearable robotics and exoskeletons, as well as Extended Reality (XR). All these CPS applications comprise of intimately coupled communication as well as compute loads, while requiring deterministic (i.e., guaranteed) "end-to-end" performance. Taking wearable robotics as an example, the interworking of Occupational Exoskeletons (OEs) allows reducing the physical load of a human worker which nevertheless imposes demanding requirements on computation and supplicated hardware [12]. Computational offloading, such as to the edge, offers significantly more efficient implementations of OEs in terms of energy consumption, efficiency and costs. However, such computational offloading necessitates a new type of support by the infrastructure in the *end-to-end context*. It is paramount here to stress the difference between the notion of *end-to-end* with respect to applications in the future cyber-physical continuum, versus its connotation traditionally in the design of networks. In the future, sufficient end-to-end performance refers to supporting the application communication and computation demands *over the entire loop*, from sensors via cyber-physical representation and back to actuation modules. This end-to-end loop might comprise different wireless and wired technologies, incorporates either local or cloudified compute resources and thus spans over various different technologies, standards as well as administrative domains. End-to-end paths through such heterogeneous network infrastructures have today unpredictable latency, reliability and availability variations which directly jeopardize the operation of CPS applications. The main challenges relate therefore to an efficient technology integration of future cellular systems, TSN and DetNet domains, OPC UA as well as edge computing resources towards a *deterministic end-to-end communication & compute* infrastructure. This will not only comprise Key Performance Indicators (KPIs) such as latency, reliability and availability but also relates to time awareness, security, privacy and trust in the novel end-to-end connotation applicable to CPSs.

*Scalability* challenges of such future infrastructures are related to the ones regarding convergence but significantly go beyond them. Beyond convergence of infrastructures, capturing CPS communication & computation loads, as well as their requirements and expressing their characteristic interaction patterns from an end-to-end perspective is key to having infrastructures adapt to CPS applications in an automated fashion. Corresponding interfaces and specifications are required. Likewise, converged end-to-end communication and compute infrastructures will have to adapt to a wide span of CPS applications in terms of their workloads and requirements, in order to enable scalability. Such an adaptation only becomes possible if converged end-to-end infrastructures become predictable in terms of their resource-KPI trade-offs while being subject to various stochastic influences. It is, in particular, the end-to-end relevance of future CPS applications that make the aspects of *vertical interfacing* as well as *predictability* extremely challenging.

These emerging challenges give an idea in which sense future 6G technology must ensure end-to-end deterministic characteristics across a multiplicity of heterogeneous technologies and application domains, including wired and wireless communication infrastructure as well as compute resources. Achieving this will require substantial extensions and harmonization among future 3GPP cellular systems, TSN, DetNet and other technologies/frameworks such as OPC UA. Beyond current standards and systems, existing research has partially addressed some of the aforementioned challenges, but novel approaches are still required to a large extent. In this paper, we argue that deterministic communication systems of today are at the initial stage of their evolution rather than at their endpoint. Despite the intense research interest from academia and industry over the last decade, leading to initial technological standards and commercial products, it is clear that substantial challenges need to be addressed to achieve scalability and convergence towards a cyber-physical continuum. Key concepts to overcome these challenges will be 6G-native evolution towards deterministic communications, wireless-friendly integration regarding TSN and DetNet, data-driven characterizations of stochastic elements, end-to-end time awareness and security concepts, as well as leveraging network digital twins towards system awareness.

Recently, several surveys have reviewed the ongoing developments towards 6G mobile networks including visions for 6G, technology enablers, use cases, etc. The key performance indicators for 6G and the limitations of 5G that form the basis of 6G vision were discussed in [13]. Research works including [14]–[17] have explored technological enablers (e.g., THz communication, edge intelligence, blockchain, swarm networks and data-driven management and operations) for 6G communication networks. In [18], five use case families have been identified for 6G by the European 6G flagship project Hexa-X. Supporting these innovative use cases in future communication networks requires an underlying deterministic communication infrastructure. In the context of deterministic communications, specification work is ongoing in different standardization organizations (i.e., IEEE, IETF and 3GPP) along with several research initiatives towards 6G. However, the requirements of end-to-end interworking and integration of different technologies (URLLC, TSN at Layer 2 and DetNet at Layer 3, edge computing systems) for deterministic communication systems have not been investigated systematically in these works. The key aspect of enabling end-to-end determinism in future communication systems is largely missing.

Given this state-of-the-art, our main contribution in this paper relates to first a comprehensive summary of state-of-the-art technologies in the field, namely 5G URLLC, TSN and DetNet, which we present in Section II. From this characterization, we discuss the newly arising challenges and recent research around these challenges in Section III. Finally, in Section IV we outline future directions of system evolution toward scalable deterministic communication systems and their application scenarios. In particular, we argue that the current discourse around 6G systems, with standardization starting at approx. 1.5 years, is a unique opportunity to address the identified and necessary system features.

## II. Status Quo of Deterministic Communications in Standards and Industrial Practice

Fundamentally, determinism refers to absolute certainty with respect to the further state evolution of a process or a system [19]. A deterministic system is therefore a system for which the next state (or sequence of states) is certain, given a current state. In cyber-physical systems, certainty in the communication and/or in the execution of a computation is desirable, nevertheless, absolute certainty does not exist [20]. Thus, approaches are required that can deal with the remaining uncertainty in a communication or computation system. The term *deterministic communications* refers to such an approach in the context of communication. In this approach stochastic requirements of a CPS application are given while the stochastic uncertainty of the communication system can be quantified [21]. It is thus possible to determine if the level of uncertainty of the communication system is acceptable regarding the application requirements of the respective CPS. The involved stochastic requirements can be arbitrarily high, turning the communication system into a pseudo-deterministic one. To determine if a communication system fulfills the requirements, typically a dependability analysis is performed [22]. As dependability analysis of a system requires arguing with respect to (various sources of) stochastic uncertainty, the system must be predictable to a certain degree, meaning that the stochastic characteristics of the system's future state evolution must be derivable in a quantitative way. These derivations can rely on bounding conditions that need to be fulfilled. Throughout this paper, the term *deterministic communications (and compute infrastructure)* refers to scenarios where the deployment of a CPS is only successful if the involved communication (and compute) system can determine quantitatively the fulfillment of the stochastic requirements of the corresponding CPS application.

Today, three standards stand out with respect to their focus on deterministic communications, namely IEEE's TSN, IETF's DetNet and 3GPP's 5G Time-Sensitive Communication (TSC) and URLLC. IEEE 802.1 Time-Sensitive Networking (TSN) plays a central role in the wired Ethernet domain, as it allows to provision guaranteed high-performance connectivity services for traffic flows on a common Layer 2 bridged Ethernet infrastructure. Traditionally, industrial networking consisted of a plethora of fieldbus technologies and their descendant real-time industrial Ethernet variants e.g., (PROFINET, EtherCAT, etc.), providing deterministic communication services. However, they tend to be incompatible with each other. In future industrial networks,

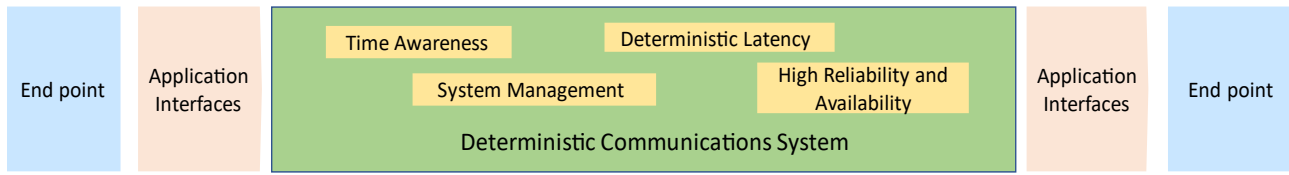
**Figure 1** Basic characteristics of deterministic communication systems.

TSN is expected to provide basic deterministic communications for various industrial applications. TSN aims to create a common standard for converged real-time communications in industrial networks [23].

As an evolution of TSN, DetNet realizes deterministic communications in Layer-3 routed networks while being a technology with a focus on wired communication infrastructures as well. DetNet has been standardized by the corresponding Working Group (WG) of the IETF since 2014. In close collaboration with the IEEE TSN TG, IETFs DetNet WG has proposed a common architecture to support deterministic services at both L2 and L3.

The flexibility of wireless communications also motivated the provisioning of deterministic communications over wireless networks. The 3rd Generation Partnership Project (3GPP) has made significant efforts to enable 5G systems to support latency-sensitive use cases through URLLC. The ambition of 5G URLLC has been to provide low-latency communication together with high reliability for maintaining strict one-way Radio Access Network (RAN) latency bounds such as 1 ms with 99.999% probability. URLLC is defined in the 5G standard Releases 15 to 17. Interworking with TSN referred to as 5G TSC has been addressed in Releases 16 and 17, and 5G support for DetNet is being prepared in Release 18.

Finally, as a further complement to the deterministic communication technologies TSN, DetNet and 5G, the OPC's Field Level Communication (FLC) is an initiative that develops one common multi-vendor middleware framework for a converged network infrastructure [23]. The OPC framework relies on either Pub/Sub or Client-Server as the architecture principle in organizing the middleware functionality. OPC leverages currently TSN (while in the future it is expected to be extended to DetNet and 5G/6G) and standardizes interoperable data models and common procedures across a variety of industrial use cases with multi-vendor components. This includes the specification and usage of deterministic communication patterns through initializing and configuring the underlying deterministic communication infrastructure.

Despite the fact that the above standards have defined to some extent the interoperability among one another, the industrial practice today mostly sees isolated operations of these technologies in individual domains/verticals. This isolated operation nevertheless limits the application range of these technologies and masks inefficiencies in their interoperability. Due to the foreseeable increase in CPS application use cases, the current state-of-the-art as well as industrial practice are rather a first step towards deterministic communications at scale.

All state-of-the-art deterministic communication systems today exhibit several characteristics that are all essential for supporting CPS application traffic. These characteristics can be grouped into five categories (shown in Figure 1):
1. Time awareness
2. Deterministic latency
3. High reliability and availability
4. System management
5. Application interfacing

In the following, we discuss how these five features are supported today by the key deterministic communication standards in the wireless domain (5G) and wired domains (TSN, DetNet). Moreover, we discuss industrial best practices in these contexts. In Section III we will then make the argument that in the future evolution, the somewhat different characteristics of the below systems need to be harmonized and extended towards full end-to-end support.

### I. Time Awareness

Time awareness is a key to configuring the operation of deterministic communications and ensuring the synchronization between the physical and digital world, for example, the coordinated movement of two robotic arms must have time awareness to synchronize the movement. In general, access to a common time reference for all nodes in the network is necessary to support many time-critical applications. First, network nodes need to be accurately and precisely synchronized with each other, e.g., to perform desirable time-triggered operations according to a globally coordinated periodic schedule. Secondly, applications running on the endpoints also need to be aware of time, e.g., to undertake certain actions at a particular time. Overall, time synchronization protocols and mechanisms exist today both in the wired (TSN) and wireless domains (5G).

#### 1) Time Synchronization in TSN and DetNet

Accurate time synchronization in packet-switched networks is often achieved using the IEEE 1588 Precision Time Protocol (PTP). In PTP, the clocks in the secondary nodes synchronize to an accurate source of time called the primary Grandmaster (GM) clock through the exchange of PTP messages [24]. Multiple PTP profiles have been specified for different use cases and environments. To enable sub-microsecond time synchronization for time-sensitive applications, e.g., in industrial networks or audio-video production, IEEE 802.1AS has been introduced [1]. IEEE 802.1 AS defines a specific profile of PTP called generalized Precision Time Protocol (gPTP), which is applicable for Ethernet transport. gPTP optimizes PTP for time-sensitive applications by constraining the values of PTP parameters (e.g., periodicity of the sync and follow-up messages) and requires additional features to accurately compensate for delay variation, e.g., by tracking and correcting the frequency offset between neighbors [25]. In addition, to support fault tolerance (link or GM failure), the standard also introduces redundancy in the form of multiple gPTP domains and redundant GMs. In contrast to TSN, DetNet does not specify any time synchronization mechanism but leverages existing mechanisms provided by lower layers, e.g., IEEE 1588 and IEEE 802.1AS.

*2) Time Synchronization in 3GPP and ITU-T*

Time synchronization is an integral part of advanced mobile network technologies such as 5G. Standard development organizations (SDOs) including 3GPP and International Telecommunication Union Telecommunication Standardization Sector (ITU-T) play key roles in determining and developing network synchronization solutions and architectures for 5G mobile communication technology and further for 6G. Time synchronization within the 4G & 5G RAN is needed for the operation of the radio access network including Time Division Duplexing (TDD), Coordinated Multipoint Transmission (CoMP) and carrier aggregation. Fundamentally, within a 5G system there is one GM clock provided typically by the Global Navigation Satellite System (GNSS) for other 5G entities to synchronize themselves. Within mobile networks, the common reference time of the 5G Grandmaster (GM) clock is distributed in the RAN via the transport network to the base stations (called gNB in 5G networks) to support several RAN operations as listed above across neighboring gNBs.

As a platform for cyber-physical communication, it is required to integrate the 5G system with time synchronization domains outside the 5G system itself. For example, it is desirable to synchronize different devices and machines to some external reference clocks over the 5G system. To this end, the internal 5G time synchronization procedures need to be extended towards the edges of the 5G system, which are the User Plane Function (UPF) as a gateway to the wired communication infrastructure and the user equipment (UE) as mobile end device or gateway to a local communication network. This time distribution mechanism is depicted in Figure 2. The 5G transport network enables the time distribution from the 5G GM clock to the gNBs and UPFs, while the radio access network time synchronization mechanism enables accurate time reference delivery over the radio interface (Uu interface according to 3GPP nomenclature) from gNB towards UEs. A detailed description of this process is given in [26]. The PTP synchronization mechanism specified by IEEE 1588, allows such distributions from the GM clock to the gNBs in the 5G system. ITU-T has developed telecom profiles with a target to distribute phase/time with +/- 1.5 microseconds timing accuracy. Specifically, there are two ITU-T profiles for the distribution of time synchronization: G.8275.1 and G.8275.2 [27], [28].

Concerning the radio access network, 3GPP starting from the Release 16 has specified a mechanism that allows UEs to receive, from gNBs, the accurate time reference information via System Information Block (SIB) or Radio Resource Control (RRC) control messages [29]. In Release 17 further enhancements have been made in the time distribution over the radio interface to compensate for the propagation delay between gNB and UEs in order to reduce the time error introduced in the time distribution.

5G mobile networks can be used to provide wireless transport in a general deterministic communication network, for example as wireless connectivity in an otherwise Ethernet/TSN-based wired industrial network. In this case, time synchronization in the wired network needs to be transported from some primary grandmaster node in the wired network through the 5G network to a secondary node in another network segment that is wirelessly connected over 5G. The gPTP messages in this case are transported from an ingress node of the 5G system (i.e., a UPF or UE) to an egress node of the 5G system (i.e., a UPF or UE) over a user plane.

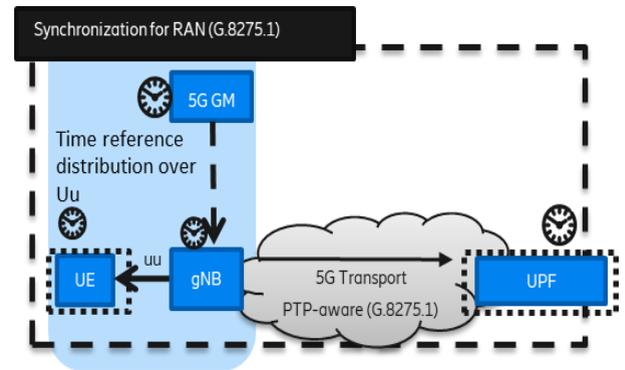

**Figure 2 Illustration of 5G's internal time-synchronization schemes.**

By time synchronizing the 5G system edges at UPF and UE to a common 5G clock (described above and depicted in Figure 2), the 5G system can determine the transit time of the PTP message through the 5G system and correct the time in the PTP message accordingly. In this way, time synchronization can be provided over the 5G network with low time error.

## II. Deterministic Latency

Traditionally, the primary performance metric in communication networks has been the achievable data rate. Latency, and to some extent the Packet Delay Variation (PDV, also referred to as jitter), have been considered for multimedia applications such as audio/video. However, the requirement on latency bounds has not been stringent. Deterministic communication technologies (5G URLLC, TSN and DetNet) aim to achieve deterministic latencies and bounded PDVs with very high confidence, which means that the focus of the requirement is on the tail mass of the systems latency distribution while the average latency only plays a minor role.

Depending on the criticality of the CPS application, the corresponding requirements on the PDV bounds can be very demanding, typically emphasizing the reliability of the communication and computation system. Each of the above-mentioned standard's technology employs different techniques to ensure the support of such real-time latency bounds in the context of CPS. Below we provide an overview of such techniques in 5G, TSN and DetNet.

*1) Deterministic Latency in 5G*
Moving from Long-term Evolution (LTE) to 5G, the cellular system capabilities enabling CPS-type real-time latency bounds are greatly improved. Starting from Release 15, several features at Medium Access Layer (MAC) and physical layer (Layer 1) as well as in the core network have been added to enable URLLC services [30]. The ambition of 5G URLLC has been to provide low-latency communication and high reliability simultaneously. The most prominent example of such ambition in 5G URLLC has been a target one-way RAN latency of 1 ms with 99.999% probability. Figure 3 provides a summary of the newly introduced radio and core network features within 5G enabling URLLC, namely (i) scalable and flexible numerology realized through the so-called New Radio (NR) Physical layer of 5G, (ii) mini-slots and short Transmission Time Intervals (TTIs), (iii) low-latency optimized dynamic Time Division Duplexing (TDD), (iv) fast Hybrid ARQ (HARQ) processing time, (v) uplink pre-scheduling with Configured Grants (CG), and (vi) robust transmission modes for control and data channels. Optimizations for the core network include redundant transmission options, and flexible selection of local breakout and edge compute location.

The result of these new features is essentially a toolbox to cater to different latency bounds by two means [31], [32]:
1. Optimizing the access time to the radio channel enabling sub-millisecond latencies (achieved by lowering processing delays, channel access delays, and others) [32],
2. Providing higher robustness of individual transmissions to achieve the same latency reliability with fewer transmission attempts, at the cost of reduced spectral efficiency due to quite conservative transmission modes defined in 5G URLLC.

*2) Deterministic Latency in TSN and DetNet*
In contrast to wireless networks, queuing delay is the key obstacle in wired networks regarding deterministic latencies. In the worst case, excessive queueing could result in packet losses due to packet dropping inside Ethernet switches. Typically, queueing delay can be bounded and packet losses due to network congestion can be avoided by allocating resources (e.g., buffers, bandwidth, transmission time) along the network path between the communicating endpoints. TSN specifies multiple traffic shaping and scheduling mechanisms in order to regulate the flow of packets in the network. IEEE 802.1 Qav defines Credit-based Shaping (CBS) where multiple flows can share link bandwidth according to the number of allocated credits [3]. Essentially, link transmission bandwidth is allocated to TSN flows. Similarly, TSN flows can also be scheduled to link transmission time, as specified in IEEE 802.1Qbv – the TSN Enhancements for Scheduled Traffic. Cyclic Queuing and Forwarding (CQF), as defined in IEEE 802.1Qch, is a specific use of scheduling where queues are serviced in an alternate fashion to provide deterministic latency per hop [33]. In addition, TSN traffic shapers/schedulers are complemented by other mechanisms specified in the TSN standards suite that supports deterministic latency for TSN flows. Frame preemption (IEEE 802.1Qbu) allows the preemption of low-priority frames by a high-priority frame, i.e., the transmission of a less urgent frame can be interrupted against an urgent frame [34]. The transmission of the interrupted frame can be resumed after the transmission of the high-priority frame is finished. Frame preemption allows for the reduction of the guard time interval between a low-priority frame and a high-priority frame, although is dependent on the transmission time corresponding to the minimum Ethernet frame size.

As of today, DetNet by itself has not yet specified congestion control or traffic shaping/scheduling mechanisms ensuring deterministic latency but assumed the support of lower layers (e.g., TSN, MPLS-TE) to support time-critical flows. Work on specific queueing mechanisms for DetNet is only starting up

| End-to-End URLLC communication services | | | |
|---|---|---|---|
| UE | gNB | Transport Network | Core Network |
| 5G Radio Access Network | | 5G Core Network | |
| - Scalable and flexible numerology<br>- Mini-slots and short TTIs<br>- Low-latency optimized dynamic TDD<br>- Fast processing time and fast HARQ<br>- Pre-scheduling on uplink with configured grants (CG) (Layer 2)<br>- Robust link adaption<br>- Robust control channels<br>- Dual connectivity | | - Support of redundant transmission on N3/N9 interface<br>- Redundant at transport layer<br>- Redundant User plane paths on multiple UEs Per device<br>- Edge computing | |

**Figure 3 Features enabling URLLC communication services in 5G.**

in IETF standardization.

*III.  High Reliability and Availability*
Latency bounds in an end-to-end context must be paired with a sufficient likelihood to successfully convey a cyber-physical interaction. Due to the varying requirements of a cyber-physical system, the sufficient level of this likelihood is intimately linked to the cyber-physical context. From a CPS application perspective, the metrics: reliability and availability, are mostly mentioned in the context of system dependability, which nevertheless can comprise the analysis of further attributes like safety, security, and maintainability [35]. Concerning communication systems, IEC 61907 defined communication dependability as the "ability to perform as and when required to meet specified communication and operational requirements" [22]. It is important to recognize that the terms *availability* and *reliability* have different definitions, stemming from the fact that these metrics originated in the real-time computing community [35]. Availability in this context refers to the fraction of time during which a communication system provides latencies as required by the application over a total duration. Note that this definition can directly be measured. On the other hand, reliability is understood as the likelihood of witnessing a future latency-bound violation under the condition that the communication system currently performing correctly. A common measure for reliability is for instance the mean time between failures. A result of these definitions is that a communication and/or computation system with a high availability does not necessarily need to have also high reliability and vice versa. Furthermore, the conversion of packet transmission reliability of a communication system into availability and reliability is more involved. Due to this, as an example, 3GPP starting from Release 16 has worked on the definition of communication service reliability and availability in specifications [36] with respect to supporting communication for automation in several vertical domains. Despite these intricacies, a system with a high or very high packet transmission reliability (i.e., a low or very low packet error rate) typically has also a high availability and reliability. Therefore, in the following discussion, we mostly focus on reliability aspects of communication systems as understood from a communications perspective, focusing again on 5G, TSN and DetNet [36].

*1) Achieving Reliability (and Availability) in 5G*
Unlike wired networks, the radio link quality in 5G is a major determining factor for reliability and latency. Depending on the Signal-to-Interference-and-Noise-Ratio (SINR) of the link, reliability measured as the block error rate can vary significantly over time. A solution for reliable wireless transmission with high spectral efficiency is to apply link adaptation together with HARQ retransmissions to recover from unsuccessful transmissions. The above-mentioned acceleration of corresponding processing times has made this feature applicable for URLLC also for stringent latency requirements. In addition, 5G has proposed mechanisms not only for data channels but also for control channels to improve reliability using robust transmission modes and various diversity techniques to boost both reliability and availability of a 5G system such as uplink multi-antenna techniques, low rate Modulation and Coding Scheme (MCS), multi-slot transmission for grant-based channel access, robust Channel Quality Information (CQI) reporting, etc, [37].

In addition to the above mechanism for extra-robust radio transmission models, reliability in 5G can be further improved by using redundancy techniques. The redundancy options in 5G differ from each other depending on the segment to which the reliability mechanism is applied. The RAN segment is typically the most unreliable part of the 5G system, its redundancy is the major determinant. Its reliability can be improved via various diversity techniques such as multiple frequency carriers, multiple antenna transmission, or multiple UEs [38]. In addition to the above enhancements, advanced interference prediction and mitigation techniques are specified that focus on managing interference from neighboring UEs and gNBs.

Additionally, the 5G core network redundancy solutions are implemented by organizing Network Function (NF) sets, each containing a group of equivalent NFs having the same functionality and sharing the same context. The NF sets can be deployed in multiple geographical locations in order to boost the total availability of the 5G system.

Despite all these technical advancements in 5G, the issue of performing a dependability analysis of a 5G system or link remains challenging due to the unquantified uncertainty in a wireless communication scenario. In other words, it is open to which extent 5G systems are predictable.

*2) Achieving Reliability (and Availability) in TSN and DetNet*
As discussed above, resource allocation combined with traffic shaping can prevent packet losses stemming from congestion in a wired (for instance Ethernet) network. However, even in such a scenario, a time-critical flow can still be disrupted if there is a failure in the network (e.g., a node or link failure) resulting in packet losses. To protect against network failures in TSN using redundant paths, IEEE 802.1CB describes a method called Frame Replication and Elimination for Reliability (FRER) [5].

FRER works by sending multiple copies of each Ethernet frame over maximally disjoint fixed network paths [39]. FRER also specifies an elimination mechanism that allows discarding the surplus packets by looking at the sequence number of packets. Frame replication and elimination can be performed both at end nodes as well as at relay nodes (intermediate Ethernet switches) depending on the network deployment and reliability requirements.

As TSN flows, DetNet flows can leverage protection against failures in the network. DetNet service protection mechanisms such as packet replication and elimination (failure protection), encoding (media error protection) and re-ordering (in-order delivery) are specified in [6]. Analogous to the FRER mechanism in TSN, DetNet specifies the Packet Replication, Elimination and Ordering Function (PREOF) to provide dedicated protection for DetNet flows against network failures. In addition, DetNet flows also benefit from fixed network paths, which can be established via explicit routes to protect against temporary outages due to the finite convergence time of routing algorithms [40].

Generally speaking, dependability analysis in wired systems is less challenging due to the better predictability of the wired medium. The remaining uncertainty from congestion can be bounded or even mitigated by applying suitable resource allocation and scheduling schemes.

*IV. System Management*

In order to achieve and maintain latency bounds, as well as reliability and availability, deterministic communication systems require several management functions. Resource management for instance guarantees the efficient use of transmission capabilities toward the latency and reliability goals defined by a CPS. Admission control ensures that the offered load in combination with the required latency and reliability pairs of the cyber-physical application supported by the deterministic communication systems does not outgrow the transmission capabilities of the system. In the following, we provide an overview of how TSN, DetNet, and 5G implement such admission control and resource management techniques.

*1) System Management in 5G*

5G systems in general, as all prior generations of cellular systems, are centralized systems featuring a control as well as a user plane. This distinction implies a set of system resources to be reserved for system control communication and thus guarantees the protection of system control messages against any temporary overload with respect to arriving user payload data. Based on this principle, all major aspects of the system management such as admission control, resource allocation and quality of service provisioning are guaranteed to be realizable in a cellular system.

System management is related to network architecture, which has evolved over the different generations of mobile networks. In 5G two different architecture options have been specified which are related to how 5G is introduced into the market and its dependency on existing 4G networks. These lead to different consequences in the context of support for deterministic communications. On the one hand, 5G systems can be realized as non-standalone or standalone architecture. Non-Standalone (NSA) relates to a system with new NR radio access technology and an existing LTE core network. Here, the NR cell acts as a booster cell. It is a typical deployment today for public wide-area mobile broadband services. The downside of NSA is nevertheless the legacy LTE control plane which leads to many extended latencies for CPS systems. In addition, NSA deployments do not support essential features for deterministic communications such as Ethernet / TSN bridging, PTP time-synchronization through 5G, etc. In contrast to NSA, standalone systems rely on the new service-based architecture (SBA) where a 5G base station is connected to a 5G core and all control and data signaling is handled by the 5G core, leading to reduced latencies. A second architecture distinction relates whenever a 5G system is operated as a public or Non-public Network (NPN). A localized 5G NPN is formed when a 5G system is deployed to enable private communication service for a dedicated organization or authority with a defined group of authorized devices, in contrast to wide-area mobile networks providing communication to the public. With respect to the use of 5G systems in the context of cyber-physical systems and industrial automation, NPN deployments are seen today as the most promising configuration [41] [42].

Furthermore, within 5G systems, there are different architectural features that allow its integration and efficient interaction with other technologies supporting resource management in the context of end-to-end deterministic communication services. Concerning support of the different types of traffic including, transport of Ethernet frames, starting from Release 15, 5G standards support Ethernet connectivity along with IP-based connectivity with Ethernet Packet Data Unit (PDU) sessions. A PDU session provides connectivity to a User Equipment (UE) towards a User Plane Function (UPF) via RAN. The Session Management Function (SMF) and Access and Mobility Management (AMF) are part of the 5G control plane which ensures the functioning of user plane connectivity. Today, 5G support for Ethernet is compatible with IEEE 802.1 standards. There are two technical enablers in 3GPP that support the integration with Ethernet-based industrial networks namely the capability of the 5G system to act as an Ethernet bridge that supports time-sensitive communication for TSN, and 5G virtual network (VN) groups [42].

Standardized exposure interfaces aim for a simplified 5G network management solution that hides the implementation complexity of the 5G system and enables simplified integration with existing application frameworks used in the context of deterministic communication services (e.g., OPC UA, ROS). 3GPP has developed features that enable seamless integration of the application frameworks on the northbound interface of the 5G system via standardized APIs. In Release 15, the focus of 3GPP was to provide simplified application programmability interfaces (APIs) that can be triggered by external applications, including functions for onboarding, discovery, secure communications, authentication, and authorization for hosting APIs. The result was a unified Common API framework (CAPIF) to expose APIs. Further, the Service Enablement Architectural Layer (SEAL) was standardized during Release 16, which allows the development and exposure of the specific functionalities across 3GPP network functions in a harmonized manner. Release 17, and 18 further introduced application-specific functionalities such as those required for smart manufacturing, automotive or public safety. Additionally, the Network Exposure Framework (NEF) has been defined by 3GPP with the purpose of providing fine-grained telecommunications APIs [43]–[45].

*2) System Management in TSN and DetNet*

For TSN networks there are three TSN configuration models proposed in the TSN standards [4], namely the fully distributed model, the fully centralized model and the centralized network distributed user model. The three models differ in the way various TSN entities interact with each other and are most relevant in the way resource management is realized in TSN, comprising a selection of network paths, allocating resources on these paths, e.g., scheduling time-triggered traffic on these paths, etc. These TSN configuration models are specified in 802.1Qcc, which also provides detailed specifications for the fully centralized configuration model, which is the only TSN configuration today with a completed standard specification.

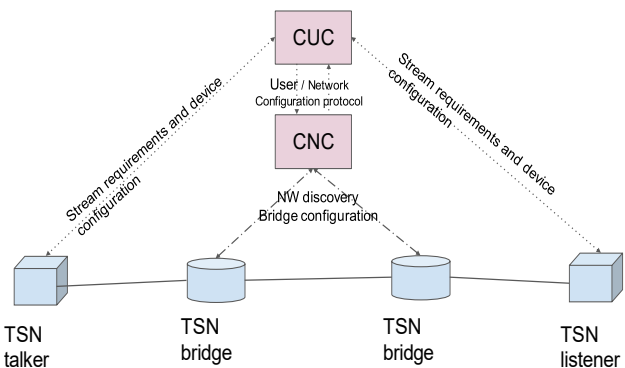

**Figure 4 Illustration of the fully centralized TSN configuration model.**

The fully centralized TSN configuration model is shown in Figure 4 [4]. The Centralized User Configuration (CUC) is an entity responsible for gathering the flow specifications and communicating them to the Centralized Network Configuration (CNC). The CNC is responsible for monitoring the network, calculating new network configurations based on the information received from the CUC or the network itself, and deploying the configuration in the network nodes.

In practice, the fully centralized TSN configuration includes the following steps:
- Traffic characteristics and service requirements are provided for individual TSN flows from the CUC to the CNC.
- The network topology, characteristics, and capabilities of TSN bridges are collected by the CNC.
- The transmission paths for TSN flows are determined by the CNC based on the service requirements, network capabilities and traffic distribution in the network.
- The TSN bridges (and potentially the TSN end-stations) on the transmission path are configured with TSN traffic shaping policy to be applied for the TSN stream along with configuring functions for priority handling, policing and filtering traffic flows, etc.

For DetNet flows, too, resource allocation in the intermediate DetNet nodes along the routed path is required to ensure that performance guarantees are met. However, scalability is a key requirement for DetNet, given a large number of flows in IP networks in contrast to a LAN with a much smaller number of TSN flows. The DetNet control plane is responsible for creating and removing DetNet flows that imply path computation, establishment, and resource allocation [40]. In addition, the DetNet control plane establishes explicit paths needed in PREOF functions. In analogy to TSN, the DetNet control plane can be of three types: fully distributed, fully centralized (Software-Defined Networking (SDN)) as well as some mixture of the two.

*V. Application Interfacing*

Finally, some functionality is required to bind into application layer frameworks, typically being domain-specific, and allowing integration of applications into a deterministic communication domain. For instance, in industrial automation, a key challenge faced before a large-scale operation is to achieve a high level of interoperability among industrial systems, which are typically sourced from a diverse set of vendors. Various efforts have been undertaken to standardize communication in industrial networks.

Over the last decade, the Open Platform Communications Unified Architecture (OPC UA) has gained a lot of traction for industrial automation solutions [23] and represents today perhaps the most important framework in industrial automation. OPC UA is specified by the OPC Foundation and ensures interoperable data exchange in industrial automation among devices and equipment from multiple vendors. Currently, there are three major efforts in the realm of OPC UA and TSN to provide deterministic communication. First, the "Field Level Communication" initiative is developing OPC UA FX – Field eXchange specification for field-level communications using both the client-server and publish-subscribe (PubSub) models including initial "application Quality of Service (QoS) requirements" definitions [46]. Second, within IEC/IEEE P60802 "TSN Profile for Industrial Automation" definitions are considered that include "Industrial Traffic Type" representations [47]. Lastly, the third effort is the IEEE P802.1Qdj "Configuration Enhancements for Time-Sensitive Networking" which includes enhancements to the CNC's northbound interface (IEEE 802.1Qcc) [48] allowing corresponding deterministic stream requests to be launched. The combination of these specifications allows for setting up end-to-end deterministic PubSub streams over TSN networks, fulfilling the application's QoS with the support of some specific traffic types.

## III. Challenges and State-of-the-Art in Research

The development of the above-mentioned technologies over the last decade has clearly advanced significantly with respect to the principal aspects of deterministic communications (time awareness, deterministic latency, reliability, system management and application interfacing). Nevertheless, with respect to several key challenges the existing industrial standards still fall short, leading to only a limited application range as of today. In contrast to today's deterministic communication systems of the first generation (i.e., defined during the last decade), future systems of the second generation will have to achieve two central goals: *scalability* as well as *convergence*. Only if the associated challenges related to these two central goals can be overcome, CPS in multiple application domains will be deployable on networked infrastructures, leading to the cyber-physical continuum.

We outline these challenges, together with the state-of-the-art in research in the following. We identify four central challenges: (I) *Predictability*, i.e., predictable, stochastic performance characterizations especially regarding the radio access network; (II) *End-to-end technology integration* concerning a more efficient integration of 5G/6G wireless systems into TSN and DetNet, as well as true end-to-end support considering edge computing systems, (III) *End-to-end security* provisioning as well as (IV) *Scalable and flexible vertical interface*s. Figure 5 illustrates the relationship between the goals of scalability and convergence, as well as the associated challenges of end-to-end technology integration, security, predictability and scalable/flexible interfacing with the applications.

*I. Predictability*

Wireless communications and networking are well known to be subject to severe communication channel distortions which are notoriously hard to predict, mitigate and/or compensate for. Despite several decades of technological advancement, still today wireless communication systems exhibit performance characteristics that are inferior to corresponding wired systems in terms of system capacity and reliability. Typical Ethernet technology today provides gigabits per second throughput with frame error rates below $10^{-6}$. In contrast, state-of-the-art wireless technology like 3GPPs 5G can achieve comparable single KPI, but not simultaneously. Furthermore, the amount of resources (in terms of bandwidth, power, diversity degree and error correction complexity) required to realize such performance goals is significantly higher in the case of wireless systems in comparison to wired ones.

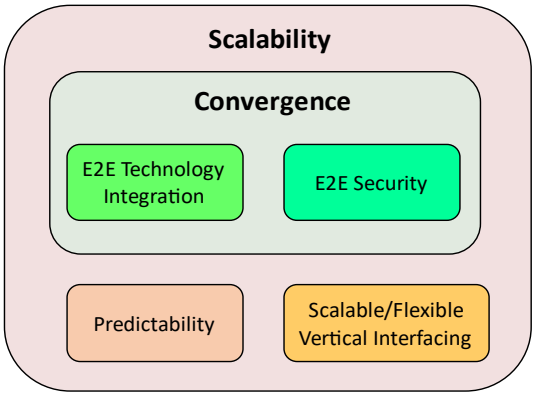

**Figure 5 Goals of the future deterministic system and the associated challenges.**

Given the significant interest in industrial automation use cases over the last decade, and the associated efforts to devise wireless communication systems for such use cases, efforts were also undertaken to experimentally determine the viability of ultra-reliable low-latency communication systems. Early efforts are reflected for instance in several system concepts and prototypes [49]–[51] showing the principal viability of URLLC-like systems. Accompanying URLLC standardization, several URLLC trials have been conducted with respect to 3GPP 5G Release 16 features. In [52], an outdoor trial was conducted leveraging the new frame structure of 5G NR with repetitions, polar coding, as well as space-frequency block coding. Small packet sizes between 50 – 200 Bytes were considered, paired with transmission distances between 300 m and 1 km, switching between stationary and mobile settings. The trials demonstrated the principal feasibility of 1 ms link layer-to-link layer latency in downlink and uplink with 99.999% packet success rate for the chosen scenarios. A second, independent study in [53] considered the comparison of 5G Enhanced Mobile Broadband (eMBB, a flavor of 5G that emphasizes achievable average rate) and URLLC implementations with respect to typical industrial automation traffic patterns and in typical indoor propagation scenarios. The trials leveraged a 3.6 GHz center frequency, as well as a 28 GHz center frequency paired by the bandwidth of 80 and 100 MHz. The study found average application layer end-to-end latencies in UL and DL to be generally between 5 and 10 ms for 5G MBB systems, whereas a latency of 10 ms was easily passed if considering the 99.9% quantile for MBB systems. In contrast, URLLC test systems showed consistent application layer end-to-end latency below 1 ms for UL and DL cases considering the average latency as well as the 99.9% quantile.

Despite the experimental evidence for ultra-reliable and low-latency communications in practice, wireless systems remain subject to various stochastic influences due to channel effects, mobility as well as various interference effects. With respect to predictability, this implies that deterministic communications in the realm of wireless communications and networks can only be achieved with respect to a stochastic guarantee. Stochastic guarantees for communication systems in general, and wireless systems in particular, have been extensively studied in the literature in the past [54]–[57]. Depending on the intent, such guarantees have been derived either from information-theoretic, queuing-theoretic or algorithmic principles. From an information-theoretic perspective, a significant stream of fundamental contributions resulted from new rate bounds for noise-limited communication channels in the finite block length regime serving theoretically as a physical layer performance guarantee in the presence of stochastic noise [54] and channel fluctuations [55]. In a second major development, significant works turned to renewed queuing-theoretic analysis, either developing novel tools to provide stochastic bounds on wireless communication systems through stochastic network calculus [56], [57] or turning to novel metrics like the Age-of-Information [58], which allow stochastic guarantees over random traffic arrival as well as service patterns [59], [60]. Finally, a third major area contributing to stochastic guarantees has evolved out of algorithmic scheduling theory, where aspects of resource allocation can be shown to result in (stochastically) guaranteed system features [61], [62]. However, a central challenge of all these approaches is their inherent embedding in mathematical modeling and subsequent derivations. It remains to date open to expand approaches towards valid stochastic guarantees that are held in practice.

## II. End-to-End Technology Integration

Typical industrial automation processes as well as cyber-physical systems consist often of closed-loop control systems, where actuation signals are periodically generated by a controller based on the input received from one or multiple sensors. This might happen at regular intervals or might be event-triggered. For example, to move a robotic arm to a desired position, the position of the arm is read by a sensor (e.g., potentiometer) and sent to the controller which generates the appropriate signal for the actuator (e.g., motor) to move the arm, as shown in Figure 6. A fundamental challenge in such systems is the fact that the CPS performance depends on the latency between the time when sensor input is generated and the time when the control signal (command) reaches the actuator. This "end-to-end" latency over the full control loop, however, does not correspond to the traditional understanding of end-to-end latency, as it comprises potentially a multitude of different links, and more importantly, it comprises compute elements that transform sensor data into actuation commands. Future deterministic communication systems will have to support such "end-to-end applications", necessitating end-to-end convergence across different deterministic communications (i.e., 5G, TSN, DetNet) and computing (e.g.,

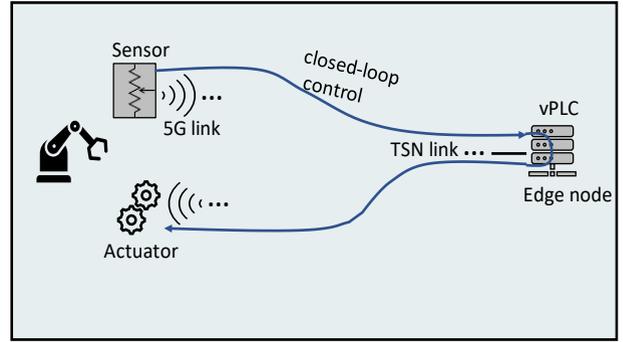

**Figure 6 Illustration of closed-loop control running on a converged TSN and edge infrastructure.**

The integration of 3GPP 5G with IEEE TSN is addressed in Release 16-17 of the 3GPP specifications, whereas the integration of 3GPP 5G with IETF DetNet is still in the early stages and specification is ongoing in Release 18 to extend the 5G's TSC framework to support integration. 3GPP in respective specifications have proposed transparent integration of 5G system with TSN and DetNet. As 5G-DetNet integration architecture is quite similar to 5G-TSN integration architecture, we will only focus on the state-of-the-art relevant to 5G-TSN integration.

Figure 7 illustrates the integration of a 5G system with a TSN network as specified by 3GPP in Release 16. The 5G system is perceived by the rest of the network as a set of TSN-capable bridges, one per UPF. For a 5G system to function as a TSN bridge, the use of TSN Translator (TT) functions for data plane traffic is introduced. The TTs provide the ports of the virtual 5G TSN bridge towards the neighboring TSN bridges. For the edge computing) technologies. Next, we discuss various integration architectures across different deterministic communication domains along with the relevant challenges.

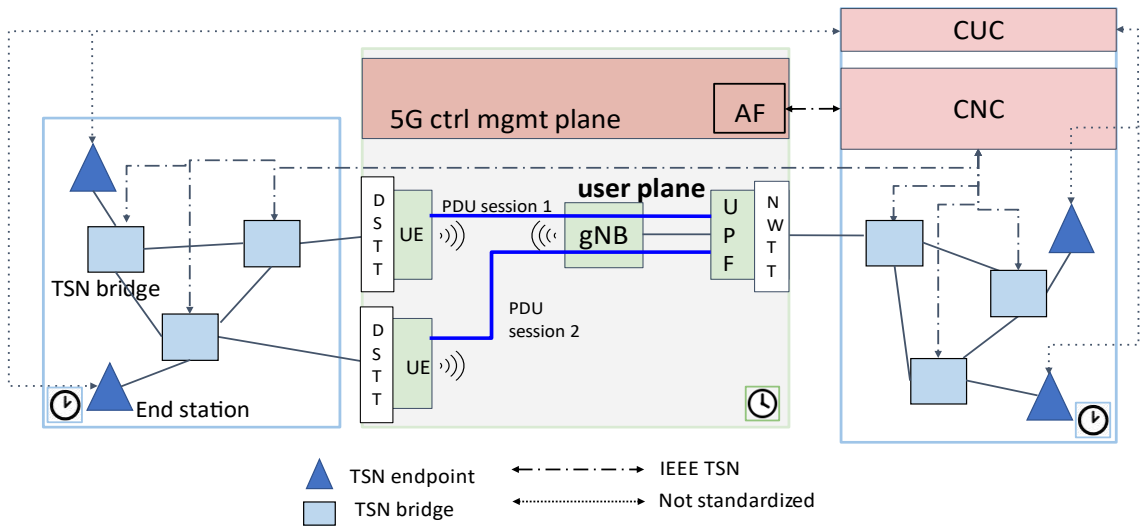

**Figure 7 Illustration of the 5G-TSN integration architecture.**

user plane, two types of TTs are defined: (i) device TT (DS-TT) on the UE side and (ii) network TT (NW-TT) on the UPF side. All UEs, the corresponding DS-TTs and the UPF form one virtual 5G system bridge. Therefore, a 5G system can be mapped to multiple virtual bridges: one per UPF. The control plane integration is achieved using TSN Application Functions (AFs). The CNC interacts with the TSN AF for the control and management of the 5G system. The TSN AF exposes the capability of the virtual TSN bridge, the network topology and port identification, etc.

The integration architecture assumes that using TTs, a sufficient level of integration can be achieved between 5G and regardless of characteristics of 5G virtual bridges vis-à-vis TSN bridges. In other words, the 5G system is perceived to a large extent as a 'wired' deterministic communication node. However, despite several optimizations specified in 5G URLLC targeting low-latency and high-reliability communication, 5G systems still exhibit significant stochastic behavior as compared to wired TSN bridges. This results in an integration that compromises either latency or reliability leading to significant challenges for end-to-end deterministic communications. For instance, end-to-end traffic scheduling in a 5G-TSN integrated network is non-trivial to implement. A time-critical packet might miss the transmission opportunity somewhere in the network path if the 5G system bridge's PDV is too high. The impact of PDV on scheduling can be compensated to some extent using the hold-and-forward buffers mechanisms at DS-TTs and NW-TTs [63]. However, the actual implementation of such mechanisms in TTs is not specified by 3GPP and is left open. It is not clear how many queues (per buffer) will be required per TT, how traffic classes will be mapped to them and how they will be scheduled.

A further, fundamental challenge arises from the pivot in cyber-physical systems toward edge computing. With Industry 4.0, a control application is decoupled from the physical platform and there has been a shift from specialized hardware-based Programmable Logic Controllers (PLCs) towards software-based controllers. Software controllers or virtual PLCs (vPLCs) are consolidated on a cloud platform in order to achieve flexibility and reduce costs. To guarantee network performance for industrial control applications (e.g., motion control, robotics), the end-to-end latency between the devices and the vPLC needs to be small and deterministic. To this end, the use of edge computing infrastructure for hosting vPLCs has been proposed. The essence of the edge computing paradigm is to perform computing tasks at the network edge, i.e., closer to the source of data in contrast to the traditional cloud computing approach where computation is done in a small number of large data centers typically further away from data sources [8]. Due to the proximity to the compute and storage, the application response times are reduced along with bandwidth savings. Therefore, by offloading industrial controllers to the edge infrastructure cost-savings can be achieved while meeting the delay requirements of closed-loop control.

The convergence of deterministic communication networks (e.g., URLLC, TSN, DetNet) with edge computing to support time-critical applications in industrial environments has been investigated recently. In [64] and [65], various 5G deployment options are considered for integration with edge computing. These works focused mainly on generic architecture aspects while integration was treated especially from a reliability perspective, but end-to-end support of determinism and time-critical applications have not been the focus. The major challenge here is to seamlessly orchestrate and configure the compute and network resources across different domains, i.e., edge computing, wired, wireless, etc., in an integrated way, such that end-to-end application requirements (e.g., latency, PDV and reliability) are met. The use of new generations of (container-based) virtualization platforms and the migration towards open and cloud-native 5G architecture compromises latency and reliability objectives in 5G systems. This hinders the achievement of seamless end-to-end deterministic communications.

Another challenge to extending the edge architecture and data plane toward supporting deterministic and reliable communication is how to ensure predictable and deterministic latency experienced by requests while traversing different software layers in a virtualized environment [66], [67]. Moreover, the current cloud management platforms (e.g., Kubernetes) have limited capability to configure the cloud execution environment (e.g., CPU scheduling, virtualized network configuration) properly considering the application requirements and TSN/DetNet configuration [65].

*III. End-to-End Security*

Ensuring a secure end-to-end flow of data between cyber-physical continuum endpoints becomes more challenging in dynamic, uncontrolled environments. It is especially true when multiple heterogeneous technologies are involved and the latency requirements are particularly vulnerable to easy to exploit Denial of Service (DoS) attacks. Today security solutions (e.g., firewalls) are not well adapted for deterministic communications. They either need to be made to work in real-time or the scheduling needs to consider the additional delay. The latter is referred to as delay transparency and it concerns delays due to access control, authorization, authentication, etc, [68]. At a high level, the following aspects need to be considered when conceiving security solutions from an end-to-end perspective:

1. The requirement for the performance of security mechanisms is not the same in the end-to-end communication paths and in the edge.
2. Besides detecting attacks on the time determinism, security mechanisms also need to ensure that packet delivery and consistency are not disrupted.
3. Applications requiring determinism have different requirements that need different security levels.
4. Time awareness mechanisms, e.g., time synchronization solutions have a big attack surface and today only isolated solutions specific to each communication technology exist [24].

Due to the strict timing requirements of TSN/DetNet, potential vulnerability points are introduced in an integrated 5G-TSN network. For instance, the need for accurate time synchronization in a TSN network makes DoS attacks very effective, e.g., a time-critical flow can be disrupted by a long PTP outage, slowDoS-type attacks can be sufficient to disrupt sensitive applications [69]. Similarly, multiple TSN flows shaped via IEEE 802.1Qbv could be impacted due to one misbehaving node (e.g., not adhering to its configured schedule). An attacker using a false identity might even masquerade as a genuine endpoint to connect to the network if the authorization mechanisms are not adequate.

Security is heavily investigated in many 5G/6G research projects (notably INSPIRE-5Gplus[1] and HEXA-X[2]) but these initiatives do not sufficiently address the end-to-end perspective for deterministic networking [70]–[73]. Furthermore, few research efforts are ongoing to develop security solutions for TSN/DetNet. In [74], the use of the CBS algorithm (specified in IEEE 802.1Qav) for protecting TSN-based car systems from DoS attacks by allowing only valid traffic patterns (e.g., analyzed using end-to-end latency and the number of frames) was investigated. Several other works considered 802.1Qci's PSFP, for instance, [75] presented a survey that compares works related to AVB and TSN security. In [76], DoS attacks were identified as the biggest hidden danger that needs to be considered in the design of TSN architectures. A PSFP-based anomaly detection system was designed and evaluated by the authors. Authors in [77] investigated a centralized solution for policy management and also combined TSN stream configuration with dynamic Secure Group Tags (SGT) to achieve end-to-end security in TSN by enabling the definition and enforcement of an access policy on all the networking devices [78]. The vulnerabilities in the IEEE 1588 PTP were studied and the need for a monitoring unit was identified that compares clock offsets/delay measurements provided by numerous secondary devices [79]. These measures only tackle a specific aspect of security and do not propose a global security-by-design architecture that encompasses end-to-end, adaptive, differentiated, multi-domain, multi-provider, multi-stakeholder security that will be needed in next-generation deterministic networks.

Standards to improve security in deterministic networking have been developed such as IEC 62243 [80] and IEEE 802.1Qci [81]. IEC 62243 introduces a functional reference model with five security levels, segmenting them into zones and conduits, and defining the main requirements for system security. Zones regroup physical and functional assets with similar security requirements, and conduits support communication between the zones. IEEE 802.1Qci introduces Per-Stream Filtering and Policing (PSFP) which detects whether a stream violates the defined behavior and takes mitigating actions accordingly. The concept of defense-in-depth is also introduced. With respect to DetNet, RFC 9055 specifies the different security mechanisms of the IETF norm where potential security threats are analyzed and mitigation is provided by path redundancy, encryption, dummy traffic insertion, integrity protection, node authentication and control message protection [82].

### IV. Scalable/Flexible Vertical Interfacing

To support a cyber-physical continuum over a heterogeneous infrastructure (e.g., TSN, 6G, DetNet), scalable interfaces are required that support dynamic and automated configurations and reconfigurations. This is challenging particularly when the infrastructure is sourced from multiple vendors and multiple communication technologies that might differ from each other in terms of their QoS mechanisms. OPC UA FX has been addressing the interoperability challenge to some extent. However, current OPC UA mechanisms towards deterministic communication exclusively relate to TSN networks, require manual configuration of devices and support only static scenarios. From an automation perspective, several extensions to OPC UA FX are planned that will alleviate bottlenecks. OPC UA FX "Offline engineered TSN" will allow for device specifications that can be incorporated during the planning phase such that engineered traffic during the planning phase can meet end-to-end QoS targets. With OPC UA FX "Plug&Produce TSN" system specifications are further enhanced to allow engineered traffic to meet end-to-end quality of service requirements based on the actual network topology and resource utilization. Thus, device configuration will be eased through these extensions from fully manual configuration during the installation phase, over configurations generation through planning tools and automatic configuration download to devices, to a fully automated reconfiguration at runtime. Taking these developments into account, still further significant steps need to be taken towards truly scalable provisioning of a cyber-physical continuum, majorly related to stochastic uncertainties. On the communication side of end-to-end systems, reconfigurations might become necessary due to changing communication characteristics. Therefore, a more comprehensive definition of allocatable application QoS levels and potentially available network guarantees are required. To this end, QoS mechanisms for deterministic communication technologies (i.e, 6G, TSN and DetNet) and their parameter mapping need to be abstracted by defining network guarantees (for individual traffic types) and related network policies (potential concurrent combinations of network guarantees). Also, it is unclear how a specific mapping of application-level QoS to network policies translates into parameter settings across a network consisting of heterogeneous communication technologies (e.g., integrated 5G/6G-TSN network). Expanding on this end-to-end view, an area entirely untouched to date is the aspect of expressing application characteristics as well as compute capabilities through corresponding interfaces/standards. Subsequent interfaces for configuration and dynamic management will be required as well.

---

[1] https://www.inspire-5gplus.eu/

[2] https://hexa-x.eu/

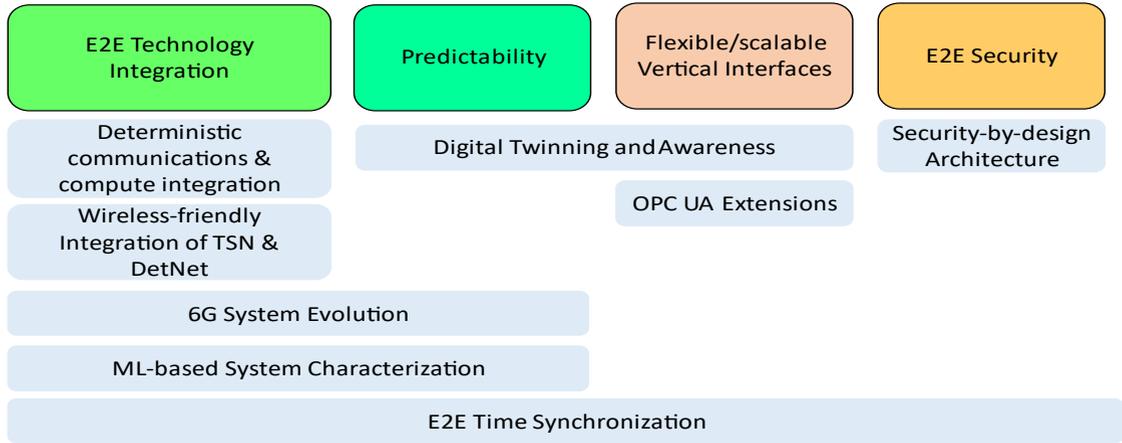

**Figure 8 Four challenges of future deterministic communication systems and corresponding enablers to address them.**

## IV. Vision toward Next-Generation Deterministic Communication Systems

While the last decade has seen a significant increase in attention towards deterministic communications in wireless and wired networks, it is clear today that these efforts must be seen as an initial stage rather than a final one. 5G URLLC, TSN and DetNet represent remarkable technological developments as of today and have been picking up commercial applications in respective domains and verticals. Nevertheless, it is also a fact that industrial adoption has been slower with respect to all three technologies than perhaps anticipated several years ago. This is despite the growing number of applications requiring deterministic communication services and support. As argued in the above section, two main goals need to be achieved in the further evolution of deterministic communications/flexible vertical interfaces. In the following, we outline potential approaches and enablers to address these challenges, namely (i) 6G system evolution; (ii) Machine Learning (ML) -based system characterization; (iii) Wireless friendly integration of TSN and DetNet; (iv) end-to-end time synchronization; (v) Digital twinning and awareness; (vi) deterministic communications and compute integration; (vii) Security by design; and (viii) OPC UA extensions. The relationship between these approaches/enablers as well as the four challenges discussed in Section III is visualized in Figure 8.

### I. 6G System Evolution

5G systems today are subject to at least two deficiencies when it comes to the support of deterministic communications on the air interface. Most importantly, the PDV remains quite high in 5G systems. Even with the introduction of URLLC, typical PDVs are in the range of milliseconds. While these variations are strongly improved in comparison to corresponding values from LTE, they are still no match in comparison to the corresponding PDVs of TSN for instance over an Ethernet connection (which is typically in the range of a few microseconds). 6G systems are likely to define features that address these challenges, requiring 6G transceivers to be built for high-precision time delivery of packets, and the entire transmission system to be revisited with respect to reducing PDVs. By this, 6G systems will enable easier integration into technologies likes TSN and DetNet. A second step in 6G systems might be the redefinition of measures to achieve URLLC characteristics with a particular focus on mechanisms that improve the predictability of wireless communications. To this end, mechanisms could be enhanced that monitor and/or contain all sources of communication uncertainty. For instance, besides inherent wireless channel uncertainties, interference is a major source of uncertainty in wireless communications [83]–[86], opening the possibility for scalable, field-deployable and predictable wireless interference control. Finally, commercialization success of URLLC today is partially precluded due to the high cost of corresponding terminal-related chipsets. 6G systems will address these issues and can be expected to reduce the involved costs while achieving URLLC performance.

### II. Wireless-friendly Integration of TSN and DetNet

Today's integration of 5G systems into TSN is implemented in a way that 5G systems assimilate to the system characteristics of TSN standards, i.e., 5G systems project themselves as 'wired' deterministic communication nodes. Similar work has been initiated within 3GPP in support of DetNet. However, despite the efforts of 3GPP to introduce more deterministic communication modes in 5G like URLLC, 5G systems are still subject to substantial sources of randomness due to their wireless nature. As a result, when integrating into TSN or DetNet today, either 5G systems need to compromise on determinism while preserving wired node-like latencies or they need to compromise on latency while preserving determinism.

Thus, a central aspect in the next generation, of converged deterministic communications will be to account for the differences in characteristics of the wired TSN bridge and the 6G virtual TSN bridges in an efficient manner. For instance, bridge capabilities exposed by a 6G system to, for instance, a TSN CNC will have to be more representative of the wireless system behavior rather than of a wired TSN bridge. Metrics like maximum and minimum delay between port pairs are thus not sufficient to capture the latency-reliability characteristics of wireless bridges. In future architectures, richer node characteristics will have to be exposed. These characteristics could be obtained via ML, as discussed above, and would include for every port-pair, e.g., the quantiles for the expected latency, the prediction horizon, confidence interval, etc. Using these node characteristics, a deterministic communications scheduler can generate end-to-end schedules for the TSN/DetNet nodes and QoS mapping for the 6G system. The focus of past research has been on the scalability of the scheduling methods or on optimizing a criterion, e.g., minimizing the number of resources (queues) used, reducing the flow-span, etc [87], [88]. However, for future wireless-wired integration, the scheduler design and control plane interactions will have to be revisited. In particular, robustness against PDV from a wireless bridge could be taken into account while generating end-to-end schedules. To this end, a more dynamic model is needed between network nodes and the CNC interactions.

A further necessity regarding wireless-friendly integration is to better accommodate for changes in these characteristics. When bridge characteristics change, a new interaction between a 6G bridge and the TSN CNC will have to be triggered, leading to a replanning of the traffic handling for all flows passing through this bridge. Fast fluctuations of wireless transmission may be covered by a larger PDV and do not require a new interaction with the CNC. However, a change in the carrier frequency of a UE (e.g., between 3.7 GHz and 26 GHz) implies a different characterization of the latency behavior for the port of the virtual bridge represented by the UE. Also, when UEs are connected to or disconnected from the UPF, new bridge ports are added or removed. All these events lead to interactions with a TSN CNC causing a possible replanning of many traffic flows. Resource allocation approaches will be required that can efficiently handle this replanning efficiently (even when faced with 100s of connections as typically being the case with 5G and certainly 6G wireless cells).

### III. ML-based System Characterization

A key success factor to future deterministic communication systems is the efficient integration of stochastic elements along an end-to-end path as they are arising in the context of cyber-physical systems, i.e., including wireless links as well as edge and cloud computing elements. It is evident that such an efficient integration can only be facilitated if the probabilistic latency/reliability nature of such elements is accounted for, in contrast to today's static link characterizations for instance in TSN which mask stochastic behavior by advocating seemingly deterministic link specifications. However, from a characterization point of view, this opens up the question of how stochastic elements can be captured in their respective features. As discussed in Section III, model-based approaches to characterize such stochastic elements fall short of being applicable to real systems. It is likely that data-driven machine learning methods will have to be leveraged in combination with corresponding architectures that facilitate efficient data collection and distribution. Machine Learning has seen over the last five years a steeply increasing interest with respect to its study in various different uses for wireless systems, for instance, with respect to resource allocation, the substitution of individual functions of transceiver structures, the substitution of entire transceiver structures in particular through variational autoencoder approaches, as well as with respect to the application layer when it comes to federated learning. In contrast to these research avenues, for future deterministic communication systems it will be necessary to (i) devise ML schemes that accurately capture tail probabilities of distributions; (ii) find light-weight approaches in terms of training data and run-time retraining for 6G systems; and (iii) derive from such insights transceiver designs that support predictability, i.e., different transceiver designs (and exposure of corresponding characteristics) will lead to different profiles in terms of ML-based predictability (see also previous Section IV.III).

### IV. Digital Twinning and Awareness

Considering wireless friendly implementation of an end-to-end deterministic communications system, some control measures must be provided to improve PDV or QoS algorithms such as scheduling, prioritization, and similar. Most importantly, the environment of the wireless communication system can allow or hinder a seamless constant quality communication which needs to be addressed. The standard approach is to use long-term statistics in specific scenarios and consider an approximation of the channel conditions depending on the channel model being considered. Of course, existing channel measurements and channel estimations are already helping to provide high communication quality. But a leap forward is essential to derive communication parameters from the environmental observations, which means predicting channel behavior based on those observations to make use of rerouting, replanning, or rescheduling ahead of a potential loss in the line of sight or similar factors. The observations can be made by a wide field of sensing for communication which helps the communication system to provide an uninterrupted communication service. Therefore, it is beneficial to have sensing technologies integrated in a 6G communication system, to enable enhanced TSN-AF and CNC interactions that plan ahead and can anticipate effects of, e.g., changing access channel condition. Obviously, standards need to be provided for different sensor technologies so that sensors from different manufacturers can provide standardized data for sensor fusion and can be integrated into the communication system with or without ML. This helps to reduce the variance of stochastic components in the wireless communication system as more and more of the time-varying changes in the environment can be considered and processed. Additionally, it is possible to use UEs themselves as sensing devices to generate coverage maps for path planning, route planning, or similar tasks. Current technology developments for 6G are allowing to influence the channel and coverage map via Reconfigurable Intelligent Surfaces (RIS) so that nearly line of sight quality can be reached if an alternative path via RIS can be used [89]. This environmental awareness can drive the network as an active component in the 6G system which provides better control of stochastic parameters in wireless systems in general for an improved communication quality.

In order to realize such awareness functions on future systems, digital twins (DTs) are seen as technical enabler. A DT is a virtual representation of real-world entities and processes [90]. It is based on models and data structures to describe the state, observations, and relations of real-world objects. By continuous data collection about real-world objects, the digital twin can maintain an up-to-date view of physical reality. Digital twinning is generally expected to play an increasing role in many fields in the future [91] and is one enabler for establishing the cyber-physical continuum [10]. Mobile networks can include sensing capabilities while positioning of the mobile user equipment is one such capability that has been supported and evolved for mobile networks for a long time. With 5G it is possible to provide e.g., high precision positioning for UEs in a factory environment [42], [92]. For 6G significantly richer sensing capabilities are envisaged, which are investigated under the concept of Joint Communication and Sensing (JCS) [91], [93], [94]. Radar-like capabilities could be provided by 6G or support for Simultaneous Localization and Mapping (SLAM). DTs can also be integrated into the 6G system (i.e., a 6G DT), to monitor, plan and optimize the network operation [95]. It can be used for a range of use cases such as communication service assurance, advanced radio network planning, and performance prediction.

Both environmental and situational awareness can be useful for the 6G network and the CPS. For instance, they can be useful for the 6G network to estimate the spatial availability of communication service to external users whereas for CPS, they can monitor surroundings of human workers in a shopfloor and predict expected behavior. CPS digital twins may in addition have other sources to create situational awareness such as the digital twin of a factory connected to a machine vision system from which environmental information can be obtained. However, environmental, and situational awareness are unexplored areas that require further investigations on the design of 6G CPS interactive architecture.

*V. End-to-End Time Synchronization*

As discussed above, future deterministic communication services will encompass heterogeneous wired-wireless infrastructures along with multiple application domains. To ensure resilient time synchronization over these heterogeneous infrastructures, the future systems should not only address large variations in synchronicity needs from various cyber-physical applications but also ensure their smooth functioning in the event of failures.

Solutions need to be designed to ensure the tight synchronicity budget of future wired-wireless 6G systems by minimizing the time between the ingress and egress of the 6G systems. In the context of 5G-TSN, the current synchronicity budget requirements according to the Release 17 specifications is 900ns, which is expected to become more stringent with the evolution of new 6G use cases.

The current standardization efforts in TSN and 3GPP focused on minimizing the time error budget and providing resilient time synchronization. The 5G time sensitive communications [96], provides procedures on how to use the 5G system as a relay or a boundary or a transparent clock for 5G-TSN. The need for a fault tolerant time synchronization is currently being addressed in IEEE P802.1ASdm – hot standby amendment [97]. The resilient time synchronization should be carefully designed to incorporate redundant GM and redundant domains to ensure minimum time synchronicity budget and provide seamless transition from primary time domain to hot standby time domain in case of failures.

Finally, most time synchronization techniques are not designed with security in mind. Hence, given the tight synchronicity demands of the 6G use cases, time synchronization is subject to a large attack surface, which is likely to increase as systems will be increasingly deployed in uncertain environments. Solutions need to design to provide secure transmission of timing messages from the E2E perspective to avoid scenarios where adversaries could delay or manipulate time exchange messages [98], [99].

*VI. Deterministic Communications and Compute Integration*

With respect to edge computing, one major need is to move towards real-time cloud computing, to give recommendations on how to design and offer deterministic edge computing services for deploying reliable and time-critical applications on the edge. This has to cover architectural, but also deployment aspects of how the edge computing domain and the deterministic communications network should be integrated by considering possible combinations of different types of 3GPP Non-Public NPNs, deterministic communication network options (TSN, DetNet) and edge computing deployment models.

Regarding resource orchestration, the aim will be to develop a framework for the federated handling of resources, covering the edge computing domain, the wireless and wired networks, and the centralized data center by considering the application requirements. Furthermore, the framework could be extended to cover a spectrum of cloud resources, including public and private cloud resources provided as hybrid cloud and also including a variety of edge computing resources of different sizes and positioned in different locations, sometimes referred to as near edge resources (on-premises/customer, regional) and far edge resources (application/storage server sets, gateways, special devices, etc). The variety of edge computing resources, local and central (datacenter) cloud resources form what can be called the edge-to-cloud resource continuum. The integration and use of such an edge-to-cloud resource continuum in the resource orchestration framework can be considered one of the key elements for building a cyber-physical continuum. Also, requirements and strategies need to be investigated for how timing guarantees in the cloud platform can be ensured by cloud resource management (e.g., CPU allocation, scheduling of virtual resources) in order to ensure deterministic operation. Furthermore, to ensure end-to-end deterministic communications (e.g., with IEEE 802.1Qbv Scheduled Traffic) some real-time user plane support functionality should be deployed in the virtualized domain, which can perform traffic scheduling according to the specified IEEE 802.1Qbv traffic rules and to support time synchronization. A further need is to cover the integration of network and cloud reliability by considering the TSN and DetNet reliability features.

Finally, the user plane, control plane and orchestrator aspects should also be considered for real-time support of edge computing, where the virtualized TSN or DetNet functions and the application instances/components have to be orchestrated in a federated way. Hence, it is required to deploy some network control plane entities into the virtualized domain, which is linked to cloud management. New interfaces would also be needed, which can allow the deep and seamless integration of the edge computing orchestration resource management system toward the control plane of the deterministic communication network segment.

## VII. Security-by-Design Architecture

In 6G deterministic communication systems, traditional network security (e.g., firewalls, IDS, etc) mechanisms may not be directly applicable because they introduce a significant amount of latency and PDV to time-critical flows.

Therefore, future communication systems need to move away from security-enhanced (patchy technologies) towards security-by-design systems where security and privacy are enabled for time-critical applications as needed. Key security concepts like trust, trustworthiness (e.g., of automation and ML techniques used), root cause analysis, liability, redundancy, etc, need to be integrated in such systems as well as end-to-end performance-awareness, high-precision telemetry, and security filtering. To support security adaptability and differentiation for different devices and grouping of streams into zones and conduits, specialized security functions and techniques are needed. Security assessment (using scalable and adaptable monitoring) of the end-to-end deterministic network chain will allow improving trust, trustworthiness, liability, resiliency, and reliability. Furthermore, detection and remediation strategies could be improved using ML techniques (e.g., Federated Learning and Multi-access Edge Computing) that use features extracted from the different networking layers and locations to enable cross layer anomaly and intrusion detection.

The security-by-design architecture needs to consider the security enablers defined in INSPIRE-5Gplus, namely for collecting the features needed, analyzing them (e.g., using AI or other techniques) and reacting to attacks. But these need to be improved and adapted to the deterministic and latency requirements. To do this, new techniques need to be introduced that include InBand Telemetry that allows for the collection and reporting of network state directly from the data plane in programmable switches (e.g., using the P4 network device programming language). Furthermore, since these techniques will negatively impact the latency, the security functions must also allow for managing different security levels and finding the best compromise between the risks and costs, including finding an optimal solution that does not need to systematically analyze every packet.

## VIII. OPC UA Extensions

As outlined in section III.IV, the next generation of interfaces for the cyber-physical continuum will have to significantly extend in terms of scalable characteristics in heterogeneous infrastructures. Within OPC UA FX, IEC/IEEE P60802 and IEEE P802.1Qdj, there are already steps planned to expand interfacing between, and mapping of, application layer and network layer to ensure automated configurations. However, going beyond that, we envision in the future the focus to be even stronger on dynamic mappings of (static as well as time-varying) application QoS requirements (or requirement ranges) to time-varying network QoS capabilities (or probability density functions thereof). Efficient technology integration as well as predictable infrastructures will enable characterizing stochastic end-to-end network capability ranges, from which applications might choose one out of several operation points. However, this requires a more stringent representation of such capabilities, while also requiring applications to provide such ranges through future interfaces. As application requirements and network capabilities evolve over time, such mappings will have to be revisited frequently, with corresponding changes to resource allocations, scheduling policies and potentially routing paths. Assuring seamless application performance while negotiating and choosing these mappings at runtime will have to be taken into consideration as well. Finally, within the same direction such dynamic mappings will have to also include the description of compute requirements as well as the possible compute capabilities at run-time for edge-cloud infrastructures. As of today, this will go significantly beyond what is within the foreseeable realm of OPC UA FX.

## V. Conclusions

Fueled by the digital transformation of societies and innovative use cases such as adaptive/mobile industrial automation, XR and wearable robotics, a convergence process is ongoing towards a cyber-physical continuum. Communication and compute infrastructures of today are not capable of efficiently integrating these upcoming use cases despite substantial emphasis on latency and reliability of communication and compute systems and infrastructures over the last decade. Instead, a fully converged end-to-end deterministic communication infrastructure is required to support the upcoming cyber-physical continuum, whereas scalable operation and service provisioning will ensure efficient implementation of the continuum. Four main challenges need to be overcome to realize such future systems: *Predictability* of stochastic communications, *end-to-end technology integration* of systems such as 5G URLLC, TSN and DetNet along with Edge computing and OPC UA, end-to-end security provisioning over heterogeneous infrastructures, as well as defining new scalable, vertical interfaces for CPS characterization towards future infrastructures. Different technology enablers exist today that have the potential to overcome these challenges, namely native 6G evolution, wireless-friendly integration of 6G into TSN and DetNet, security by design, digital twinning, data-driven characterization of stochastic system features, as well as more efficient integration of edge computing into communication infrastructures. Today, there is a window of opportunity to pave the way towards leveraging these enablers in future 6G networks.

## ACKNOWLEDGMENT

This work was supported by the European Commission through the H2020 project DETERMINISTIC6G (Grant Agreement no. 101096504).

## REFERENCES

[1] 'IEEE Standard for Local and Metropolitan Area Networks–Timing and Synchronization for Time-Sensitive Applications'. Jun. 2020.
[2] 'IEEE Standard for Local and metropolitan area networks – Bridges and Bridged Networks - Amendment 25: Enhancements for Scheduled Traffic'. Mar. 2016.
[3] 'IEEE Standard for Local and Metropolitan Area Networks - Virtual Bridged Local Area Networks Amendment 12: Forwarding and Queuing Enhancements for Time-Sensitive Streams'. Jan. 2010.
[4] 'IEEE Standard for Local and Metropolitan Area Networks–Bridges and Bridged Networks – Amendment 31: Stream Reservation Protocol (SRP) Enhancements and Performance Improvements', *IEEE Std 8021Qcc-2018 Amend. IEEE Std 8021Q-2018 Amend. IEEE Std 8021Qcp-2018*, pp. 1–208, Oct. 2018, doi: 10.1109/IEEESTD.2018.8514112.
[5] 'IEEE Standard for Local and metropolitan area networks–Frame Replication and Elimination for Reliability'. Oct. 2017.
[6] N. Finn, P. Thubert, B. Varga, and J. Farkas, 'Deterministic Networking Architecture', Internet Engineering Task Force, Request for Comments RFC 8655, Oct. 2019. doi: 10.17487/RFC8655.
[7] B. Briscoe, K. De Schepper, M. Bagnulo, and G. White, 'Low Latency, Low Loss, and Scalable Throughput (L4S) Internet Service: Architecture', RFC Editor, RFC9330, Jan. 2023. doi: 10.17487/RFC9330.
[8] W. Z. Khan, E. Ahmed, S. Hakak, I. Yaqoob, and A. Ahmed, 'Edge computing: A survey', *Future Gener. Comput. Syst.*, vol. 97, pp. 219–235, Aug. 2019, doi: 10.1016/j.future.2019.02.050.
[9] K. Zhang, Y. Shi, S. Karnouskos, T. Sauter, H. Fang, and A. W. Colombo, 'Advancements in Industrial Cyber-Physical Systems: An Overview and Perspectives', *IEEE Trans. Ind. Inform.*, vol. 19, no. 1, pp. 716–729, Jan. 2023, doi: 10.1109/TII.2022.3199481.
[10] '6G – Connecting a cyber-physical world'. https://www.ericsson.com/en/reports-and-papers/white-papers/a-research-outlook-towards-6g (accessed Dec. 11, 2022).
[11] 'Future network trends driving universal metaverse mobility'. https://www.ericsson.com/en/reports-and-papers/ericsson-technology-review/articles/technology-trends-2022 (accessed Dec. 11, 2022).
[12] 'Occupational Exoskeletons: Overview of Their Benefits and Limitations in Preventing Work-Related Musculoskeletal Disorders'. https://www.tandfonline.com/doi/epdf/10.1080/24725838.2019.1638331?needAccess=true&role=button (accessed Dec. 11, 2022).
[13] F. Tariq, M. Khandaker, K.-K. Wong, M. Imran, M. Bennis, and M. Debbah, 'A Speculative Study on 6G'. arXiv, Aug. 06, 2019. Accessed: Feb. 07, 2023. [Online]. Available: http://arxiv.org/abs/1902.06700
[14] C. Yeh, G. Do Jo, Y.-J. Ko, and K. Kyu Chung, 'Perspectives on 6G wireless communications', *Elsevier*, 2021, doi: 10.1016/j.icte.2021.12.017.
[15] W. Saad, M. Bennis, and M. Chen, 'A Vision of 6G Wireless Systems: Applications, Trends, Technologies, and Open Research Problems', *IEEE Netw.*, vol. 34, no. 3, pp. 134–142, May 2020, doi: 10.1109/MNET.001.1900287.
[16] S. Dang, O. Amin, B. Shihada, and M.-S. Alouini, 'What should 6G be?', *Nat. Electron.*, vol. 3, no. 1, Art. no. 1, Jan. 2020, doi: 10.1038/s41928-019-0355-6.
[17] C. D. Alwis *et al.*, 'Survey on 6G Frontiers: Trends, Applications, Requirements, Technologies and Future Research', *IEEE Open J. Commun. Soc.*, vol. 2, pp. 836–886, 2021, doi: 10.1109/OJCOMS.2021.3071496.
[18] M. A. Uusitalo *et al.*, '6G Vision, Value, Use Cases and Technologies From European 6G Flagship Project Hexa-X', *IEEE Access*, vol. 9, pp. 160004–160020, 2021, doi: 10.1109/ACCESS.2021.3130030.

[19] 'Determinism', *Merriam-Webster.com Dictionary*. Jan. 03, 2023. Accessed: Feb. 27, 2023. [Online]. Available: https://www.merriam-webster.com/dictionary/determinism

[20] J. A. Stankovic and K. Ramamritham, 'What is predictability for real-time systems?', *Real-Time Syst.*, vol. 2, no. 4, pp. 247–254, Nov. 1990, doi: 10.1007/BF01995673.

[21] '5G; Service requirements for cyber-physical control applications in vertical domains'. May 2022. [Online]. Available: https://www.etsi.org/deliver/etsi_ts/122100_122199/122104/17.07.00_60/ts_122104v170700p.pdf

[22] 'IEC 61907:2009 Communication network dependability engineering'. https://webstore.iec.ch/publication/6088 (accessed Jan. 09, 2023).

[23] D. Bruckner et al., 'An Introduction to OPC UA TSN for Industrial Communication Systems', *Proc. IEEE*, vol. 107, no. 6, pp. 1121–1131, Jun. 2019, doi: 10.1109/JPROC.2018.2888703.

[24] 'IEEE Standard for a Precision Clock Synchronization Protocol for Networked Measurement and Control Systems', *IEEE Std 1588-2019 Revis. OfIEEE Std 1588-2008*, pp. 1–499, Jun. 2020, doi: 10.1109/IEEESTD.2020.9120376.

[25] G. M. Garner and H. Ryu, 'Synchronization of audio/video bridging networks using IEEE 802.1AS', *IEEE Commun. Mag.*, vol. 49, no. 2, pp. 140–147, Feb. 2011, doi: 10.1109/MCOM.2011.5706322.

[26] I. Godor et al., 'A Look Inside 5G Standards to Support Time Synchronization for Smart Manufacturing', *IEEE Commun. Stand. Mag.*, vol. 4, no. 3, pp. 14–21, Sep. 2020, doi: 10.1109/MCOMSTD.001.2000010.

[27] 'G.8275.1 : Precision time protocol telecom profile for phase/time synchronization with full timing support from the network'. https://www.itu.int/rec/T-REC-G.8275.1/en (accessed Jan. 09, 2023).

[28] 'G.8275.2 : Precision time protocol telecom profile for phase/time synchronization with partial timing support from the network'. https://www.itu.int/rec/T-REC-G.8275.2/_page.print (accessed Jan. 09, 2023).

[29] D. Patel, J. Diachina, S. Ruffini, M. De Andrade, J. Sachs, and D. P. Venmani, 'Time error analysis of 5G time synchronization solutions for time aware industrial networks', in *2021 IEEE International Symposium on Precision Clock Synchronization for Measurement, Control, and Communication (ISPCS)*, Oct. 2021, pp. 1–6. doi: 10.1109/ISPCS49990.2021.9615318.

[30] O. Liberg, M. Sundberg, E. Wang, J. Bergman, J. Sachs, and G. Wikström, *Cellular Internet of Things: From Massive Deployments to Critical 5G Applications*, 2nd edition. Amsterdam: Academic Press, 2019.

[31] F. Alriksson, L. Boström, J. Sachs, Y.-P. E. Wang, and A. Zaidi, 'Critical IoT connectivity Ideal for Time-Critical Communications', *Ericsson Technol. Rev.*, vol. 2020, no. 6, pp. 2–13, Jun. 2020, doi: 10.23919/ETR.2020.9905508.

[32] '5G-SMART D1.5 Evaluation of radio network deployment options'. Aug. 18, 2022. Accessed: Jan. 02, 2023. [Online]. Available: https://5gsmart.eu/wp-content/uploads/5G-SMART-D1.5-v1.0.pdf

[33] 'IEEE Standard for Local and metropolitan area networks--Bridges and Bridged Networks--Amendment 29: Cyclic Queuing and Forwarding'. IEEE, 2017.

[34] 'IEEE Standard for Local and metropolitan area networks – Bridges and Bridged Networks – Amendment 26: Frame Preemption'. Aug. 2016.

[35] A. Gnad, 'Aspects of Dependability Assessment in ZDKI'. Jun. 2017. [Online]. Available: https://industrial-radio-lab.eu/en/publications-en/

[36] 'Service requirements for the 5G system (3GPP TS 22.261 version 16.14.0 Release 16)'. ETSI, Apr. 2021.

[37] F. Hamidi-Sepehr et al., '5G URLLC: Evolution of High-Performance Wireless Networking for Industrial Automation', *IEEE Commun. Stand. Mag.*, vol. 5, no. 2, pp. 132–140, Jun. 2021, doi: 10.1109/MCOMSTD.001.2000035.

[38] 'Integration of 5G with Time-Sensitive Networking for Industrial Communications – 5G-ACIA'. https://5g-acia.org/whitepapers/integration-of-5g-with-time-sensitive-networking-for-industrial-communications/ (accessed Nov. 25, 2022).

[39] 'IEEE Standard for Local and metropolitan area networks—Bridges and Bridged Networks - Amendment 24: Path Control and Reservation', *IEEE Std 8021Qca-2015 Amend. IEEE Std 8021Q-2014 Amend. IEEE Std 8021Qcd-2015 IEEE Std 8021Q-2014Cor 1-2015*, pp. 1–120, Mar. 2016, doi: 10.1109/IEEESTD.2016.7434544.

[40] A. G. Malis, X. Geng, M. Chen, F. Qin, and B. Varga, 'Deterministic Networking (DetNet) Controller Plane Framework', Internet Engineering Task Force, Internet Draft draft-ietf-detnet-controller-plane-framework-02, Jun. 2022. Accessed: Dec. 17, 2022. [Online]. Available: https://datatracker.ietf.org/doc/draft-ietf-detnet-controller-plane-framework

[41] '5G Non-Public Networks for Industrial Scenarios'. 5G Acia, Jul. 2019. [Online]. Available: https://5g-acia.org/wp-content/uploads/2021/04/WP_5G_NPN_2019_01.pdf

[42] '5G-Smart: Deliverable 5.2, First Report on 5G Network Architecture Options and Assessments'. [Online]. Available: https://5gsmart.eu/wp-content/uploads/5G-SMART-D5.2-v1.0.pdf

[43] G. Seres et al., 'Creating programmable 5G systems for the Industrial IoT', *Ericsson Technol. Rev.*, vol. 2022, no. 10, pp. 2–12, Oct. 2022, doi: 10.23919/ETR.2022.9934828.

[44] 'Exposure of 5G Capabilities for Connected Industries and Automation Applications – 5G-ACIA'. https://5g-acia.org/whitepapers/exposure-of-5g-capabilities-for-connected-industries-and-automation-applications-2/ (accessed Jan. 27, 2023).

[45] '5G-Smart: Deliverable 5.5, Report Describing the Framework for 5G System and Network Management Functions'. [Online]. Available: https://5gsmart.eu/wp-content/uploads/5G-SMART-D5.5-v1.0.pdf

[46] 'UAFX Part 80: Overview and Concepts'. https://reference.opcfoundation.org/UAFX/Part80/v100/docs/ (accessed Mar. 21, 2023).

[47] 'IEC/IEEE 60802 TSN Profile for Industrial Automation |'. https://1.ieee802.org/tsn/iec-ieee-60802/ (accessed Dec. 17, 2022).

[48] 'P802.1Qdj – Configuration Enhancements for Time-Sensitive Networking |'. https://1.ieee802.org/tsn/802-1qdj/ (accessed Dec. 17, 2022).

[49] V. Narasimha Swamy et al., 'Real-Time Cooperative Communication for Automation Over Wireless', *IEEE Trans. Wirel. Commun.*, vol. 16, no. 11, pp. 7168–7183, Nov. 2017, doi: 10.1109/TWC.2017.2741485.

[50] M. Serror, S. Vaaßen, K. Wehrle, and J. Gross, 'Practical Evaluation of Cooperative Communication for Ultra-Reliability and Low-Latency', in *2018 IEEE 19th International Symposium on 'A World of Wireless, Mobile*

and Multimedia Networks' (WoWMoM)*, Jun. 2018, pp. 14–15. doi: 10.1109/WoWMoM.2018.8449807.

[51] C. Dombrowski and J. Gross, 'EchoRing: A Low-Latency, Reliable Token-Passing MAC Protocol for Wireless Industrial Networks', in *Proceedings of European Wireless 2015; 21th European Wireless Conference*, May 2015, pp. 1–8.

[52] M. Iwabuchi *et al.*, '5G Field Experimental Trials on URLLC Using New Frame Structure', in *2017 IEEE Globecom Workshops (GC Wkshps)*, Dec. 2017, pp. 1–6. doi: 10.1109/GLOCOMW.2017.8269130.

[53] J. Ansari *et al.*, 'Performance of 5G Trials for Industrial Automation', *Electronics*, vol. 11, no. 3, p. 412, Jan. 2022, doi: 10.3390/electronics11030412.

[54] Y. Polyanskiy, H. V. Poor, and S. Verdu, 'Channel Coding Rate in the Finite Blocklength Regime', *IEEE Trans. Inf. Theory*, vol. 56, no. 5, pp. 2307–2359, May 2010, doi: 10.1109/TIT.2010.2043769.

[55] G. Durisi, T. Koch, and P. Popovski, 'Toward Massive, Ultrareliable, and Low-Latency Wireless Communication With Short Packets', *Proc. IEEE*, vol. 104, no. 9, pp. 1711–1726, Sep. 2016, doi: 10.1109/JPROC.2016.2537298.

[56] H. Al-Zubaidy, J. Liebeherr, and A. Burchard, 'Network-Layer Performance Analysis of Multihop Fading Channels', *IEEEACM Trans. Netw.*, vol. 24, no. 1, pp. 204–217, Feb. 2016, doi: 10.1109/TNET.2014.2360675.

[57] S. Schiessl, H. Al-Zubaidy, M. Skoglund, and J. Gross, 'Delay Performance of Wireless Communications With Imperfect CSI and Finite-Length Coding', *IEEE Trans. Commun.*, vol. 66, no. 12, pp. 6527–6541, Dec. 2018, doi: 10.1109/TCOMM.2018.2860000.

[58] S. Kaul, R. Yates, and M. Gruteser, 'Real-time status: How often should one update?', in *2012 Proceedings IEEE INFOCOM*, Orlando, FL, USA, Mar. 2012, pp. 2731–2735. doi: 10.1109/INFCOM.2012.6195689.

[59] J. P. Champati, H. Al-Zubaidy, and J. Gross, 'Statistical Guarantee Optimization for AoI in Single-Hop and Two-Hop FCFS Systems With Periodic Arrivals', *IEEE Trans. Commun.*, vol. 69, no. 1, pp. 365–381, Jan. 2021, doi: 10.1109/TCOMM.2020.3027877.

[60] J. P. Champati, H. Al-Zubaidy, and J. Gross, 'On the Distribution of AoI for the GI/GI/1/1 and GI/GI/1/2 Systems: Exact Expressions and Bounds', in *IEEE INFOCOM 2019 - IEEE Conference on Computer Communications*, Apr. 2019, pp. 37–45. doi: 10.1109/INFOCOM.2019.8737474.

[61] Y. Chen, H. Zhang, N. Fisher, L. Y. Wang, and G. Yin, 'Probabilistic Per-Packet Real-Time Guarantees for Wireless Networked Sensing and Control', *IEEE Trans. Ind. Inform.*, vol. 14, no. 5, pp. 2133–2145, May 2018, doi: 10.1109/TII.2018.2795567.

[62] S. Jošilo and G. Dán, 'Computation Offloading Scheduling for Periodic Tasks in Mobile Edge Computing', *IEEEACM Trans. Netw.*, vol. 28, no. 2, pp. 667–680, Apr. 2020, doi: 10.1109/TNET.2020.2968209.

[63] '5G-TSN integration meets networking requirements for industrial automation'. Ericsson Technology Review, 2019. [Online]. Available: https://www.ericsson.com/4ac66a/assets/local/reports-papers/ericsson-technology-review/docs/2019/5g-tsn-integration-for-industrial-automation.pdf

[64] '5G Smart: Deliverable 5.4, Second Report on 5G Network Architecture Options and Assessments'. [Online]. Available: https://5gsmart.eu/wp-content/uploads/5G-SMART-D5.4-v1.0.pdf

[65] J. Harmatos and M. Maliosz, 'Architecture Integration of 5G Networks and Time-Sensitive Networking with Edge Computing for Smart Manufacturing', *Electronics*, vol. 10, no. 24, Art. no. 24, Jan. 2021, doi: 10.3390/electronics10243085.

[66] R. Morabito, J. Kjällman, and M. Komu, 'Hypervisors vs. Lightweight Virtualization: A Performance Comparison', in *2015 IEEE International Conference on Cloud Engineering*, Mar. 2015, pp. 386–393. doi: 10.1109/IC2E.2015.74.

[67] Á. Kovács, 'Comparison of different Linux containers', in *2017 40th International Conference on Telecommunications and Signal Processing (TSP)*, Jul. 2017, pp. 47–51. doi: 10.1109/TSP.2017.8075934.

[68] O. Kleineberg and R. Hummen, 'Cyber Security for Time-Sensitive Networking (TSN) in Modern Automation Networks', *PCNE.EU - Processing and Control News Europe*, May 29, 2017. https://www.pcne.eu/article/cyber-security-for-time-sensitive-networking-tsn-in-modern-automation-networks/ (accessed Feb. 27, 2023).

[69] N. Garcia, T. Alcaniz, A. González-Vidal, J. B. Bernabe, D. Rivera, and A. Skarmeta, 'Distributed real-time SlowDoS attacks detection over encrypted traffic using Artificial Intelligence', *J. Netw. Comput. Appl.*, vol. 173, p. 102871, Jan. 2021, doi: 10.1016/j.jnca.2020.102871.

[70] J. Ortiz *et al.*, 'Intelligent security and pervasive trust for 5G and beyond networks', in *Proceedings of the 15th International Conference on Availability, Reliability and Security*, Virtual Event Ireland, Aug. 2020, pp. 1–10. doi: 10.1145/3407023.3409219.

[71] C. Benzaid *et al.*, 'Intelligent Security Architecture for 5G and Beyond Networks', Nov. 2020, doi: 10.5281/zenodo.4288658.

[72] 'Hexa-x: Deliverable 1.1, 6G Vision, use cases and key societal values'.

[73] B. M. Khorsandi *et al.*, '6G E2E Architecture Framework with Sustainability and Security Considerations', in *2022 IEEE Globecom Workshops (GC Wkshps)*, Dec. 2022, pp. 1–6. doi: 10.1109/GCWkshps56602.2022.10008585.

[74] P. Meyer, T. Häckel, F. Korf, and T. C. Schmidt, 'DoS Protection through Credit Based Metering -- Simulation-Based Evaluation for Time-Sensitive Networking in Cars'. arXiv, Oct. 21, 2019. doi: 10.48550/arXiv.1908.09646.

[75] L. Deng, G. Xie, H. Liu, Y. Han, R. Li, and K. Li, 'A Survey of Real-Time Ethernet Modeling and Design Methodologies: From AVB to TSN', *ACM Comput. Surv.*, vol. 55, no. 2, pp. 1–36, Mar. 2023, doi: 10.1145/3487330.

[76] F. Luo, B. Wang, Z. Fang, Z. Yang, and Y. Jiang, 'Security Analysis of the TSN Backbone Architecture and Anomaly Detection System Design Based on IEEE 802.1Qci', *Secur. Commun. Netw.*, vol. 2021, pp. 1–17, Sep. 2021, doi: 10.1155/2021/6902138.

[77] R. Barton, M. Seewald, and J. Henry, 'Management of IEEE 802.1Qci Security Policies for Time Sensitive Networks (TSN)', 2018.

[78] M. Seewald and R. Barton, 'INTEGRATED SECURITY-TAGGING FOR DETERMINISTIC ETHERNET', 2019.

[79] W. Alghamd and M. Schukat, 'A Detection Model Against Precision Time Protocol Attacks', in *2020 3rd International Conference on Computer Applications & Information Security (ICCAIS)*, Mar. 2020, pp. 1–3. doi: 10.1109/ICCAIS48893.2020.9096742.

[80] 'Security aspects of 5G for Industrial networks'. 5G Acia, Feb. 01, 2021. [Online]. Available: https://5g-

acia.org/whitepapers/security-aspects-of-5g-for-industrial-networks/
[81] 'P802.1Qci – Per-Stream Filtering and Policing |'. https://1.ieee802.org/tsn/802-1qci/ (accessed Dec. 17, 2022).
[82] E. Grossman, T. Mizrahi, and A. J. Hacker, 'Deterministic Networking (DetNet) Security Considerations', Internet Engineering Task Force, Request for Comments RFC 9055, Jun. 2021. doi: 10.17487/RFC9055.
[83] H. Zhang et al., 'Scheduling With Predictable Link Reliability for Wireless Networked Control', IEEE Trans. Wirel. Commun., vol. 16, no. 9, pp. 6135–6150, Sep. 2017, doi: 10.1109/TWC.2017.2719686.
[84] Y. Xie, H. Zhang, and P. Ren, 'Unified Scheduling for Predictable Communication Reliability in Industrial Cellular Networks', in 2018 IEEE International Conference on Industrial Internet (ICII), Oct. 2018, pp. 129–138. doi: 10.1109/ICII.2018.00022.
[85] C. Li, H. Zhang, T. Zhang, J. Rao, L. Y. Wang, and G. Yin, 'Cyber-Physical Scheduling for Predictable Reliability of Inter-Vehicle Communications', IEEE Trans. Veh. Technol., vol. 69, no. 4, pp. 4192–4206, Apr. 2020, doi: 10.1109/TVT.2020.2968591.
[86] T. Zhang, H. Zhang, and Z. Meng, 'Interference and Coverage Analysis of mmWave Inter-Vehicle Broadcast with Directional Antennas', in ICC 2022 - IEEE International Conference on Communications, May 2022, pp. 273–278. doi: 10.1109/ICC45855.2022.9838709.
[87] F. Dürr and N. G. Nayak, 'No-wait Packet Scheduling for IEEE Time-sensitive Networks (TSN)', in Proceedings of the 24th International Conference on Real-Time Networks and Systems - RTNS '16, Brest, France, 2016, pp. 203–212. doi: 10.1145/2997465.2997494.
[88] S. S. Craciunas, R. S. Oliver, M. Chmelík, and W. Steiner, 'Scheduling Real-Time Communication in IEEE 802.1Qbv Time Sensitive Networks', in Proceedings of the 24th International Conference on Real-Time Networks and Systems, New York, NY, USA, Oct. 2016, pp. 183–192. doi: 10.1145/2997465.2997470.
[89] M. Di Renzo et al., 'Smart Radio Environments Empowered by Reconfigurable Intelligent Surfaces: How It Works, State of Research, and The Road Ahead', IEEE J. Sel. Areas Commun., vol. 38, no. 11, pp. 2450–2525, Nov. 2020, doi: 10.1109/JSAC.2020.3007211.
[90] 'Hexa-X: Deliverable 1.3, Targets and requirements for 6G - initial E2E architecture'. [Online]. Available: https://hexa-x.eu/wp-content/uploads/2022/03/Hexa-X_D1.3.pdf
[91] H. Andersson Y, 'Joint communication and sensing in 6G networks'. https://www.ericsson.com/en/blog/2021/10/joint-sensing-and-communication-6g (accessed Jan. 02, 2023).
[92] '5G-Smart: Deliverable 5.3, Second Report on New Technological Features to Be Supported by 5G Standardization and Their Implementation Impact'. [Online]. Available: https://5gsmart.eu/wp-content/uploads/5G-SMART-D5.3-v1.0.pdf
[93] T. Wild, V. Braun, and H. Viswanathan, 'Joint Design of Communication and Sensing for Beyond 5G and 6G Systems', IEEE Access, vol. 9, pp. 30845–30857, 2021, doi: 10.1109/ACCESS.2021.3059488.
[94] 'Hexa-X: Deliverable 3.1, Localisation and sensing use cases and gap analysis'. [Online]. Available: https://hexa-x.eu/wp-content/uploads/2022/02/Hexa-X_D3.1_v1.4.pdf
[95] C. De Lima et al., 'Convergent Communication, Sensing and Localization in 6G Systems: An Overview of Technologies, Opportunities and Challenges', IEEE Access, vol. 9, pp. 26902–26925, 2021, doi: 10.1109/ACCESS.2021.3053486.
[96] 'System architecture for the 5G System (5GS) (3GPP TS 23.501 version 16.6.0 Release 16)'. 2020. [Online]. Available: https://www.etsi.org/deliver/etsi_ts/123500_123599/123501/16.06.00_60/ts_123501v160600p.pdf
[97] 'Draft Standard for Local and Metropolitan Area Networks: Timing and Synchronization for Time-Sensitive Applications — Amendment: Hot Standby'. Accessed: Jan. 09, 2023. [Online]. Available: https://1.ieee802.org/tsn/802-1asdm/
[98] L. Narula and T. E. Humphreys, 'Requirements for Secure Clock Synchronization', IEEE J. Sel. Top. Signal Process., vol. 12, no. 4, pp. 749–762, Aug. 2018, doi: 10.1109/JSTSP.2018.2835772.
[99] X. Du, M. Guizani, Y. Xiao, and H.-H. Chen, 'Secure and Efficient Time Synchronization in Heterogeneous Sensor Networks', IEEE Trans. Veh. Technol., vol. 57, no. 4, pp. 2387–2394, Jul. 2008, doi: 10.1109/TVT.2007.912327.